\documentclass[10pt]{article}

\usepackage{amsxtra,amssymb,amsfonts,amsthm,amsmath,latexsym,stmaryrd}
\usepackage{diagrams}

\theoremstyle{plain}
\newtheorem{definition}{Definition}
\newtheorem{theorem}{Theorem}

\newtheorem*{theorem*}{Theorem}
\newtheorem{lemma}{Lemma}
\newtheorem{corollary}{Corollary}
\newtheorem*{remark*}{Remark}
\newtheorem{example}{Example}

\newcommand{\refT}[1]{Theorem~\ref{T:#1}}
\newcommand{\refS}[1]{Section~\ref{#1}}
\newcommand{\refL}[1]{Lemma~\ref{L:#1}}
\newcommand{\refC}[1]{Corollary~\ref{C:#1}}

\newcommand{\refD}[1]{Definition~\ref{D:#1}}
\newcommand{\refE}[1]{Example~\ref{E:#1}}

\def\ve{{\varepsilon}}

\def\lra{\longrightarrow}
\def\hra{\hookrightarrow}

\def\ovs{\overset}
\def\lrhu{\rightharpoonup}

\def\pr{\mathop{\rm pr}\nolimits}
\def\R{{\mathbb R}}
\def\T{{\mathbb T}}

\def\Z{{\mathbb Z}}
\def\C{{\mathbb C}}

\def\H{{\mathbb H}}
\def\E{{\mathcal E}}
\def\O{{\mathcal O}}

\def\g{{\mathfrak g}}
\def\h{{\mathfrak h}}

\def\su{\mathfrak{su}}
\def\uu{{\mathfrak u}}

\def\m{{\mathfrak m}}
\def\g{{\mathfrak g}}
\def\h{{\mathfrak h}}

\def\Sum{\mathop\Sigma}
\def\i{\imath}
\def\pfi{\varphi}

\def\oH{\buildrel\circ\over H}
\def\oH1{\buildrel\circ\over H\kern-.02in{}^1}
\def\hookuparrow{{\cup\kern-.04in{}^\uparrow}}

\def\vert{\Vert}

\def\b{\mathbf b}

\def\Im{\mathop{\rm Im}\nolimits}

\def\const{\mathop{\rm const}\nolimits}

\def\Ker{\mathop{\rm Ker}\nolimits}

\def\End{\mathop{\rm End}\nolimits}

\def\pr{\mathop{\rm pr}\nolimits}
\def\Ad{\mathop{\rm Ad}\nolimits}

\def\Aut{\mathop{\rm Aut}\nolimits}

\def\id{\mathop{\rm id}\nolimits}
\def\pt{\mathop{\rm pt}\nolimits}
\def\CP{\C{\bf P}}

\def\tr{\mathop{\rm tr}\nolimits}
\renewcommand{\Re}{\mathop{\rm Re}\nolimits}

\def\bee{\begin{equation}}
\def\eee{\end{equation}}
\def\be{\begin{equation*}}
\def\ee{\end{equation*}}
\def\bal{\begin{aligned}}
\def\eal{\end{aligned}}

\begin{document}

\title{Gauge theory of Faddeev-Skyrme functionals}

\author{Sergiy Koshkin\\
Department of Mathematics,\\
Northwestern University, Evanston, IL 60208 USA\\
Email: \texttt{koshkin@math.northwestern.edu}}

\date{}



\maketitle

\begin{abstract}
We study geometric variational problems for a class of nonlinear $\sigma$-models in quantum field theory. Mathematically, one needs to minimize an energy functional on homotopy classes of maps from closed 3-manifolds into compact homogeneous spaces $G/H$. The  minimizers are known as Hopfions and exhibit localized knot-like structure. Our main results include proving existence of Hopfions as finite energy Sobolev maps in each (generalized) homotopy class when the target space is a symmetric space. For more general spaces we obtain a weaker result on existence of minimizers in each 2-homotopy class.

Our approach is based on representing maps into $G/H$ by equivalence classes of flat connections. The equivalence is given by gauge symmetry on pullbacks of $G\to G/H$ bundles. We work out a gauge calculus for connections under this symmetry, and use it to eliminate non-compactness from the minimization problem by fixing the gauge.

\end{abstract}

\section*{Introduction}

Difficulties arising in quantum field theory led some physicists to consider effective models that describe low-energy behavior of elementary particles. In {\it nonlinear $\sigma$-models} physical fields are represented by maps into homogeneous spaces $G/H$, reflecting breaking of gauge symmetry from a large Lie group $G$ to a subgroup $H$ \cite{BMSS}. Quantum particles are then described classically as topological solitons, stationary points of effective energy functionals with distinctive topology \cite{MS}. First model of this kind was introduced by Skyrme in 1961 to describe strong interactions in terms of mesonic fields. The fields are maps $\R^3\to SU_2\simeq S^3$ with $\R^3$ effectively compactified into $S^3$ by a constancy condition at infinity. Energy minimizers called Skyrmions are localized formations with point-like cores representing baryons. The degree of a map is identified with the number of baryons in a formation. Interest in Skyrme type models was recently revived in the context of holographic duality and technicolor theories \cite{NSK,Th}, where more general Lie groups naturally appear.

In this paper we are interested in a related class of models originated by Faddeev in 1975 \cite{Fd1,Fd2}. In Faddeev's case the target manifold is $S^2=SU_2/U_1$ and the energy is defined by restricting the Skyrme energy to the $S^2$--valued maps via the equatorial embedding $S^2\hra S^3$. As in the case of maps $S^3\to S^3$ whose homotopy class is characterized by a single number, homotopy classes of maps $S^3\to S^2$ are given by their Hopf invariants and the minimizers were termed {\it Hopfions}. The cores of Hopfions were expected to be interlocked circles, twisted and knotted, in contrast to pointlike cores of Skyrmions. This remained a hunch until 1997, when Faddeev and Niemi used computer modeling to show that Hopfions do have knot-like structure \cite{FN1}; their result was later confirmed by more extensive computations in \cite{BS1}. Hopfions can be lifted to stationary points of the Skyrme functional and provide insight into low-energy behavior of quantized $SU_2$ Yang-Mills theory. A recent Faddeev-Niemi conjecture generalizes these ideas to the $SU_n$ case \cite{Ch,FN2}.

Some of the functionals encountered in nonlinear $\sigma$-models are reviewed in \refS{S1}. We are able to unify most of them using the notion of coisotropy form of a homogeneous space. Although it is impossible to define a canonical $\g$--valued left--invariant form on $G/H$ it is possible to define a left--equivariant one, at least when $G$ admits a bi-invariant Riemannian metric. This is our {\it coisotropy form} $\omega^\perp$. Left equivariance is exactly the property that the right-invariant Maurer-Cartan form $dg\,g^{-1}$ has, and $\omega^\perp$ reduces to it when $H$ is trivial. For maps $\psi:M\to G/H$ the functionals can be written uniformly as 
\begin{equation*}
E(\psi)=\int_M\frac12|\psi^*\omega^\perp|^2\,
+\,\frac14|\psi^*(\omega^\perp\wedge\omega^\perp)|^2\;dm\,.
\end{equation*}
As a result, a unified treatment of Faddeev-Skyrme models becomes possible. The second term in the energy density known as the {\it Skyrme term}, is the one responsible for existence of topologically stable minimizers.

Our main focus in this paper are the topological constraints imposed on maps and their relation to the energy functional. Therefore, we consider maps defined on closed $3$-manifolds $M$ as in \cite{AK1,AK2} to avoid effects at infinity that may split composite Hopfions into simpler parts as in \cite{LY2}. For a general $3$-manifold, homotopy classes are no longer described by a single number, two levels of invariants appear instead. The primary invariant describes $2$-homotopy classes and the secondary one, defined separately within each $2$-homotopy class, classifies homotopy. The secondary invariant is a generalization of the Hopf invariant to maps into simply connected homogeneous spaces, hence we retain the name {\it Hopfion} for the energy minimizers. Our treatment of topological constraints relies on the previous work \cite{K} that recast these invariants in a form suitable for accomodating discontinuous Sobolev maps, see also \cite{AK3}.

Gauge roots of the problem are manifest in our treatment. Although one starts from maps into $G/H$, they are naturally represented by equivalence classes of flat connections on $M\times G$. This follows the original idea of Skyrme developed in \cite{AK2} for the Faddeev model. The equivalence relation is a gauge symmetry on a subbundle of $M\times G$ obtained by pulling back $G\to G/H$, we call it a {\it coset bundle}. The {\it generalized Hopf invariant} of a map becomes the Chern-Simons invariant of the representing connection, and $\omega^\perp\wedge\omega^\perp$ is essentially the curvature of the coset bundle. 

Thus, gauge theory strings together maps, the energy functional and topological constraints of the problem. Our existence theory for Hopfions is predicated on this gauge interpretation and we use gauge-fixing at a key juncture of the proof. We also undertake a detailed study of connections on coset bundles, which is of independent interest. It is tempting to speculate that our gauge interpretation retraces the intrinsic structure of the original quantum field theories, but we do not pursue this point of view here.

Analytically, Faddeev-Skyrme functionals are a particular case of {\it polyconvex} functionals common in non-linear elasticity \cite{BlM}, as was pointed out by Manton \cite{MS}. Polyconvexity is a necessary condition for existence of sufficiently regular minimizers which explains why it appears in models with classically stable soliton solutions.
Topological constraints present a new challenge not found in elasticity that mostly studies maps with contractible codomains. Without the Skyrme term we get the classical problem for harmonic maps that suffers from {\it bubbling}, topological trivialization of limits to minimizing sequences \cite{GMS1}. As in non-linear elasticity, regularity of Hopfions is a difficult issue that we do not address in this paper.

Let us also point out that in the case of the Faddeev model $\omega^\perp\wedge\omega^\perp$ is essentially the volume form of $S^2$ and $\psi^*(\omega^\perp\wedge\omega^\perp)$ is the Hodge dual to a divergence-free field on $M$. From this point of view, the generalized Hopf invariant represents the {\it helicity} of this field, and the minimization problem is a familiar one of minimizing energy under fixed helicity \cite{CDG}. Our situation can be seen as a non-Abelian generalization of this problem.

The paper is organized as follows. In \refS{S1} we review various functionals of Faddeev-Skyrme type found in the literature, and show how most of them can be rewritten using the coisotropy form. \refS{S2} describes the minimization problem for Faddeev-Skyrme functionals and gives an informal outline of ideas used to solve it. We describe the approach of \cite{AK1,AK2} based on rephrasing the problem in terms of flat connections and review some issues arising in this context. \refS{S2.5} studies elementary properties of coisotropy forms and computes the one for $S^2$ explicitly. In \refS{S3} we develop a gauge calculus for smooth connections on coset bundles, including their description in terms of {\it untwisted potentials} and formulas for gauge action and curvature. \refS{S4} contains more technical developments. We introduce Sobolev spaces of maps and connections suitable for our minimization problem, and show that topological invariants are weakly continuous under natural topology in these spaces. Our spaces include some but not all $W^{1,2}$ maps with finite Faddeev-Skyrme energy. This restriction allows us to define $2$-homotopy and {\it homotopy sectors} (classes) for them in a topologically reasonable way. Finally, in \refS{S5} we prove our main results: existence of Hopfions in each $2$-homotopy sector, and when the codomain is a Riemannian symmetric space, in each homotopy sector. A discussion of open problems concludes the paper.

\section{Faddeev-Skyrme functionals}\label{S1}

In this section we review some effective energy functionals encountered in effective models of quantum physics. All of them share the same basic structure first suggested by Skyrme and Faddeev. We then show that they can be rewritten uniformly by using coisotropy forms of homogeneous spaces.

The fields of the original Skyrme model are maps from $\R^3$ into $S^3$, where the
$3$--sphere is interpreted as the group $SU_2$ of unimodular unitary
complex $2\times2$ matrices, and only maps converging to the identity
matrix at infinity are considered. Skyrme's idea was to add to the
standard Dirichlet energy $E_2(\psi):= \frac{1}{2} \int_{\R^3} |d\psi |^2 dx$
an additional stabilizing term $ E_4(\psi):= \frac{1}{4}\int_{\R^3} |d\psi\wedge d\psi|^2 dx,$
that would prevent stationary fields from being singular as it
happens for harmonic maps \cite{GMS1}. Here the derivative $d\psi$ takes values
in the corresponding matrix Lie algebra $\su_2$ and the wedge
product $d\psi\wedge d\psi:=\Sum_{i<j} \frac{\partial\psi}{\partial
x_i}\frac{\partial\psi}{\partial x_j} dx^i\wedge dx^j$ is defined
using matrix multiplication. Because of the condition at
infinity the maps $\psi$ can be identified via the stereographic
projection with maps from $S^3$ to $S^3$ and one can talk about
their topological degree \cite{BT, DFN}. This degree serves as a constraint when
minimizing the Skyrme functional
\begin{equation}\label{e0.1}
E(\psi)=\int_{\R^3} \frac12 |d\psi|^2\, +\,\frac14 |d\psi\wedge
d\psi|^2\;dx\,.
\end{equation}
Without a constraint constant maps are obviously the only absolute
minimizers.

The Skyrme model was later generalized to maps from $\R^3$
into $G$, where $G$ is a compact semisimple Lie group \cite{DFN}.
The functional has the form
\bee\label{e0.85}
E(u)=\int\limits_M\frac{1}{2}|\psi^{-1}d\psi|^2+\frac{1}{4}|\psi^{-1}d\psi\wedge
\psi^{-1}d\psi|^2\,dm.
\eee 
with a bi-invariant metric $|\cdot|$ on $G$. For $G=SU_2\simeq S^3$ it reduces to (\ref{e0.1}) since $|d \psi|=|\psi^{-1}d\psi|$ and $|d\psi\wedge d\psi|=|\psi^{-1}d\psi\wedge\psi^{-1}d\psi|$. Being topologically stable the minimizers were expected to also be dynamically stable, i.e. behave like solitons \cite{MS}.

Another type of models emerges if one considers maps $\R^3\overset
{\psi}\lra G/H $ into the coset space of $G$ by a closed subgroup
$H$. The first model of this kind introduced by Faddeev has $G/H=SU_2/U_1\simeq S^2$, and one
can define energy by restricting (\ref{e0.1}) to the
$S^2$--valued maps via the equatorial embedding $S^2\hra S^3$. Assume $S^2\hra\R^3$ as the unit sphere so $d\psi$ is $\R^3$--valued. Then the functional of the Faddeev model can then be written as \cite{AK2}
\bee\label{e0.86}
E(\psi)=\int\limits_M\frac{1}{2}|d\psi|^2+\frac{1}{4}|d\psi\times
d\psi|^2\,dm.
\eee
Here $\xi\times\eta$ is the cross-product of two vectors in $\R^3$.

In the original formulation of the Faddeev model the functional \eqref{e0.86} was written as
\bee\label{e0.88}
E(\psi)=\int\limits_M\frac{1}{2}|d\psi|^2+\frac{1}{4}|\psi^*\Omega|^2\,dm,
\eee
where $\Omega$ is the volume form of $S^2$. Since $S^2$ is
$2$-dimensional its volume form is also a symplectic form and \eqref{e0.88}
can be generalized to $M\ovs{\psi}\lra N$ with any symplectic codomain $N$. However, physical applications led to a stronger functional introduced by Faddeev and Niemi in \cite{FN2} for maps to complex flag manifolds $X=SU_N/\T$, namely
\bee\label{e0.87}
E(\psi)=\int\limits_M\frac{1}{2}|d\psi|^2+\frac{1}{4}\sum_i|\psi^*\Omega_i|^2\,dm.
\eee
Here $\Omega_i$ form an orthobasis in the $\dim\,X$- dimensional space of invariant symplectic forms on $X$ (see \cite{Ar} for details).

Note that in all examples we have a sum of the Dirichlet term with the square-norm of an expression quadratic and antisymmetric in first derivatives, symbolically $d\psi\wedge d\psi$. Manton suggested to interpret it in \eqref{e0.1} simply as an element of $\psi^*TX\otimes\psi^*TX$ for general Riemannian manifolds $X$ as
codomains \cite{MS}. However, Manton's functional does not coincide with
the usual Skyrme functional \eqref{e0.85} for Lie groups except when $G=SU_2$, nor does it give the energy \eqref{e0.87} of the Faddeev-Niemi model except when $X=S^2$.

There is however a natural generalization of \eqref{e0.85},\eqref{e0.87} that works for arbitrary homogeneous
spaces. To describe it we introduce a Lie algebra valued $2$-form on $X=G/H$ that serves as its Maurer-Cartan form. On a Lie group one has two canonical forms, the left--invariant one $g^{-1}dg$ and the right-invariant one $dg\,g^{-1}$. Note that the latter although not invariant under the left action, is however left $\Ad_*$--equivariant, i.e. $L_{\gamma*}(dg\,g^{-1})=\Ad_*(\gamma)dg\,g^{-1}$. On a homogeneous space $G/H$ we only have left action of the group $G$. Although it is impossible to define a meaningful $\g$--valued left--invariant form on $G/H$ it is possible to define a left--equivariant one at least when $G$ admits a bi-invariant Riemannian metric (e.g. when $G$ is Abelian, compact or semisimple \cite{BtD}). 

Let $\h^\perp$ be the orthogonal complement to the Lie algebra of $H$ with respect to the invariant metric on $\g$. One can check that the form $g\pr_{\h^\perp}(g^{-1}\,d g)g^{-1}$ is horizontal and invariant under the left action of $H$ on $G$ and therefore descends to a $\g$--valued form $\omega^\perp$ on $G/H$. This form is our coisotropy form. 
Although we do not reflect it in the notation $\omega^\bot$ depends
on a choice of presentation $X=G/H$ and a choice of a bi-invariant
metric on $G$. Obviously, when $H$ is trivial $\omega^\perp$ reduces to the right-invariant Maurer-Cartan form $dg\,g^{-1}$ on $G$ and $d\psi\,\psi^{-1}=\psi^*(dg\,g^{-1})$. 

The coset space $X=G/H$ inherits a metric from $G$ by the Riemann
quotient construction. Bi-invariance implies that $(\cdot ,\cdot)_X$ is invariant under the
left action of $G$ on $X$ and $|d\psi\,\psi^{-1}|=|\psi^{-1}d\psi|$ and $|d\psi\,\psi^{-1}\wedge d\psi\,\psi^{-1}|=|\psi^{-1}d\psi\wedge\psi^{-1}d\psi|$. Therefore, if for a map $M\overset{\psi}{\lra}G/H$ we define 
\begin{equation}\label{e0.6}
E(\psi)=\int_M\frac12|\psi^*\omega^\perp|^2\,
+\,\frac14|\psi^*(\omega^\perp\wedge\omega^\perp)|^2\;dm\,.
\end{equation}
it will turn into \eqref{e0.85} for Lie groups.

For the Faddeev-Niemi energy \eqref{e0.87} the situation is slightly different. One can show that  $\sum_i|\psi^*\Omega_i|^2=|\pr_{\h}(\omega^\bot\wedge\omega^\bot)|^2$ for an orthobasis $\Omega_i$ of invariant symplectic forms on $SU_N/\T$. Therefore, we have to modify \eqref{e0.6} into
\begin{equation}\label{e0.7}
E(\psi)=\int_M\frac12|\psi^*\omega^\perp|^2\,
+\,\frac14|\psi^*\pr_{\h}(\omega^\perp\wedge\omega^\perp)|^2\;dm\,.
\end{equation}
We refer to both \eqref{e0.6}, \eqref{e0.7} as {\it Faddeev-Skyrme functionals}, although we are primarily interested in \eqref{e0.6} in this paper.

\section{Maps as connections}\label{S2}

This section outlines our approach to minimizing Faddeev-Skyrme functionals under topological constraints. The exposition is meant to describe the main ideas and glosses over subtle analytic details. Appropriate spaces of maps are introduced in \refS{S4}, where technical issues are also fully addressed.

Skyrme and Faddeev sought to minimize energy among continuous maps on $\R^3$ constant at infinity and having a given degree or Hopf invariant respectively (see \cite{Es3}, \cite{LY2} for a mathematical treatment). Instead, as in \cite{AK1,AK2} we will consider maps defined on closed $3$-manifolds to avoid dealing with effects at infinity or a boundary. Since the domain may now have nontrivial topology the topological constraint has to be modified. The degree and the Hopf invariant classify {\it homotopy classes} of maps $S^3\to S^3$ and $S^3\to S^2$ respectively. An apppropriate generalization is to minimize energy in a given homotopy class and our maps will map into simply connected homogeneous spaces. So provisionally the problem at hand is
\begin{quote}
\noindent {\bf Faddeev-Skyrme variational problem}

{\it Find a minimizer of the Faddeev-Skyrme energy (Hopfion) in every homotopy class of maps $M\to G/H$, where $M$ is a closed $3$-manifold and $G/H$ is a compact simply connected homogeneous space of a Lie group $G$ with a closed subgroup $H$.}
\end{quote}
Homogeneous spaces may admit different representations by cosets. It will be convenient to choose a coset representation in which {\it $G$ is compact, connected and simply connected and $H<G$ is closed and connected}. This can be done without loss of generality for any compact simply connected homogeneous $X$, see \cite{K}. For example, $S^2=SO_3/SO_2$ is not a good representation because $SO_3$ is not simply connected, but $S^2=SU_2/U_1$ is. We assume in the rest of the paper that such a representation $X=G/H$ has been fixed and $X$ is equipped with a metric descending from a bi-invariant metric on $G$. Then all expressions in \eqref{e0.6}, \eqref{e0.7} are well-defined. 

Classically, Faddeev-Skyrme functionals make sense only for maps that are at least $C^1$. But spaces of differentiable maps lack compactness properties convenient in variational problems and we will need to use Sobolev maps. A traditional way of defining Sobolev maps between Riemannian manifolds is the following
(see e.g. \cite{Wh,HL1,HL2}). Let $X$ be a Riemannian manifold and $X\hra\R^n$
an isometric embedding into a Euclidian space of large dimension.
Then the spaces $W^{k,p}(M,\R^n)$ are defined in the usual way and one sets
\bee\label{e0.89}
W^{k,p}(M,N):=\{\psi\in W^{k,p}(M,\R^n)|\psi(m)\in N\ a.e.\}.
\eee
But now one faces a problem of defining homotopy classes for Sobolev maps. In general for $W^{1,p}(M,N)$ maps such a notion was introduced by White \cite{Wh}, but his $n$-homotopy classes are defined
only if $[p]>\dim X$ ($[\cdot]$ is the integral part), which excludes almost all homogeneous $X$. For maps with finite Faddeev-Skyrme energy additional regularity comes not from integrability of higher derivatives but from integrability of $2$--determinants of the first derivatives. We need a version of homotopy classes that takes advantage of this regularity information. A description of homotopy classes for continuous maps $M\to G/H$ that generalizes to finite energy Sobolev maps was obtained in \cite{K} and we recall it here.

It is proved in \cite{K} that if $\psi$ and $\pfi$ are homotopic then there exists a map into the group $M\ovs{u}\lra G$ such that $\psi=u\pfi$. More precisely, $\psi(m)=u(m)\pfi(m)$ and on the left we mean the action by an element of the group $u(m)$ on $\pfi(m)\in G/H$. It is essential that $X=G/H$ be a good coset representation as above for this to hold. Since $G$ is simply connected and $\pi_2(G)=0$ for any Lie group one has $\pi_3(G)\simeq H_3(G,\Z)$ by the Hurewicz theorem.
Let $\b_G\in H^3(G,\pi_3(G))$ denote the {\it Hurewicz class}  of $G$, i.e. the one that corresponds to every homology $3$--cycle in $G$ its image in $\pi_3(G)$ under the Hurewicz isomorphism. One can express homotopy equivalence of $\psi$ and $\pfi$ in terms of the pullback $u^*\b_G$. Of course, if $\psi=\pfi$ then $u=1$ and $u^*\b_G=0$, but in general it is not necessary that the pullback vanish for $\psi$ and $\pfi$ to be homotopic. In fact, there are maps $M\overset{w}{\lra}G$ with $w^*\b_G\neq0$ but $w\pfi=\pfi$.  Consider the subgroup generated by such maps:
\begin{equation}\label{e0.8}
\O_\pfi:=\{w^*\b_G\mid w\pfi=\pfi\}<H^3(M,\pi_3(G)).
\end{equation}
We have the following 
\begin{theorem}[\cite{K}]\label{T:02}
Two continuous maps $M\overset{\psi,\pfi}{\lra}X$ are $2$-homotopic (have homotopic restrictions to a $2$-skeleton of $M$) if and only if $\psi=u\pfi$ for a continuous $M\overset{u}{\lra}G$. They are homotopic if and only if in addition
$u^*\b_G\in\O_\pfi$. The group $\O_\pfi$ only depends on the $2$-homotopy class of $\pfi$.
\end{theorem}
Note that this is a direct generalization of Hopf's homotopy classification of maps $S^3\to S^2=SU_2/U_1$. In this case any map can be lifted to $SU_2$, i.e. $\psi=u\pfi$ with a constant map $\pfi$. Now $S^3\overset{u}{\lra}SU_2=S^3$ has a well-defined {\it degree} that can be computed by pulling back the fundamental class of $S^3$ \cite{BT}. This is exactly the class $\b_G$ if we identify $\pi_3(SU_2)\simeq\Z$. One of the definitions of the Hopf invariant is as the degree of the lift $u$ and our $u^*\b_G$ is a {\it generalized Hopf invariant}. Finally, since $\pfi=\const$ the subgroup $\O_\pfi$ is trivial and two maps are homotopic if and only if the Hopf invariant vanishes.

A way to compute the Hopf invariant that works for some Sobolev maps is to pick a DeRham representative of the fundamental class and integrate it over $S^3$. Correspondingly, we will need a DeRham representative for the Hurewicz class $\b_G$. Here is its description derived in \cite{AK1}. If $G$ is a simple group then $H^3(M,\pi_3(G))\simeq\Z$ and
$\b_G$ is represented by an integral real-valued form $\Theta$ on $G$. Namely,
$$
\Theta:=c_G\tr(g^{-1}dg\wedge g^{-1}dg \wedge g^{-1}dg ),
$$
where $c_G$ are numerical coefficients computed in
\cite{AK1} for every simple group. Thus,
\begin{equation}\label{e0.9}
u^*\Theta=c_G\tr(u^{-1}du\wedge u^{-1}du \wedge u^{-1}du).
\end{equation}
In general, if $G$ is compact and simply connected then
$G=G_1\times\dots\times G_N$, where $G_k$ are simple groups. Since
$\pi_3(G)=\pi_3(G_1)\oplus\dots\oplus\pi_3(G_N)\simeq\Z^N$:
$$
H^3(M,\pi_3(G))\simeq
H^3(M,\Z)\otimes\pi_3(G)\simeq\Z\otimes\Z^N\simeq\Z^N
$$
and we identify $H^3(M,\pi_3(G))$ with $\Z^N$. Therefore $\b_G$ is
represented by an integral vector-valued form $\Theta:=(\Theta_{\g_1},\dots,\Theta_{\g_N})$, where
\bee\label{e0.11}
\Theta_{\g_k}:=c_{G_k}\tr(\pr_{\g_k}(g^{-1}dg)\wedge\pr_{\g_k}(g^{-1}dg)\wedge\pr_{\g_k}(g^{-1}dg))
\eee
and $\g_k$ are the Lie algebras of $G_k$. Accordingly, $\O_\pfi$ from
(\ref{e0.8}) becomes a subgroup of $\Z^N$ that we denote by the same symbol. Now we can handle Sobolev maps by picking a smooth reference map $\pfi$ and allowing $u$ to be a Sobolev map. {\it A map $u\pfi$ can be declared homotopic to $u\pfi$ when $u^*\Theta$ is integrable and $\int_M u^*\Theta\in\O_\pfi$.}

Appearance of $a:=u^{-1}du$ in both the functional \eqref{e0.85} and the topological constraint suggests that it plays a  role in the variational problem. This is a $\g$-valued $1$-form on $M$, i.e. a section of $M\times\g$, where $\g$ is the Lie algebra of $G$. Already Skyrme suggested interpreting it as a connection. It is indeed the {\it gauge potential} of a connection on $M\times G$, see \cite{AK1}. We may fix $\pfi$ once and for all since all maps homotopic to it are of the form $u\pfi$. Then we wish to think of $a$ as representing $\psi=u\pfi$. However, there is a problem since non-trivial maps may satisfy $w\pfi=\pfi$. In other words, we need an equivalence relation on potentials. To do so introduce the {\it isotropy subbundles}:
\begin{equation}\label{e0.10}
\begin{aligned}
H_\pfi & :=\{(m,\gamma)\in M\times G\mid\pfi(m)=gH,\ g^{-1}\gamma g\in H\}\subset M\times G,\\
\h_\pfi & :=\{(m,\xi)\in M\times\g\mid\pfi(m)=gH,\ g^{-1}\xi
g\in\h\}\subset M\times\g.
\end{aligned}
\end{equation}
If we identify maps from $M$ to $G$ with sections of $M\times G$ then sections of $H_\pfi$ are exactly the maps with $w\pfi=\pfi$. Bi-invariant metric on $G$ induces an orthogonal decomposition of $M\times\g=\h_\pfi\oplus\h_\pfi^\bot$ and the corresponding decomposition of $\g$-valued forms. The subalgebra condition for $\h$ and invariance of the metric imply
\bee\label{shh}
[\h,\h]\subset\h\quad ,\quad [\h,\h^\bot]\subset\h^\bot,
\eee
and therefore
\bee\label{shhfi}
[\h_\pfi,\h_\pfi]\subset\h_\pfi\quad ,\quad [\h_\pfi,\h_\pfi^\bot]\subset\h_\pfi^\bot.
\eee
\begin{definition}[Isotropy decomposition of gauge potentials]\label{D:isotrop}
Let $a\in\Gamma(M\times\g)$ then its isotropic and coisotropic components are respectively
\bee\label{e0.49}
a^\vert:=\pr_{\h_\pfi}(a),\quad a^\bot:=\pr_{\h_\pfi^\bot}(a).
\eee 
\end{definition}
We will see in \refS{S3} section that it is only the coisotropic component $a^\bot$ that actually represents $\psi$. Moreover, if we set
$$
D_\pfi a:=a^\perp+\pfi^*\omega^\perp
$$ 
then the Faddeev-Skyrme energy (\ref{e0.6}) of $\psi=u\pfi$ becomes
\begin{equation}\label{e0.13}
E_\pfi(a)=\int_M\frac12|D_\pfi a|^2\, +\,\frac14|D_\pfi a\wedge
D_\pfi a|^2\;dm.
\end{equation}
In addition, $u^*\Theta$ in (\ref{e0.9}) also has a very simple
expression in terms of $a$:
\begin{equation}\label{e0.14}
u^*\Theta=c_G\tr(a\wedge a\wedge a).
\end{equation}
Thus, we succeeded at reformulating both the functional and the constraint of our variational problem in terms of the potential $a$. The advantage of this point of view is two-fold. First, it is easier to deal with a linear space of potentials rather than non-linear space of maps into $G$. Second and most important, in the course of minimizing the functional we will have to address the inherent ambiguity in the choice of $u$ to claim any kind of compactness for minimizing sequences. In terms of $a$ this ambiguity is reflected in the fact that $E_\pfi$ in \eqref{e0.13} does not depend on the isotropic component $a^\vert$, which therefore is not controlled by it. This difficulty is common in gauge theory and can be resolved by a standard technique known as {\it fixing the gauge} \cite{MM}. It turns out that $a^\vert$ itself can be interpreted as a potential of a connection, albeit in a somewhat non-traditional way, on a subbundle of $M\times G$ obtained by pulling back $G\to G/H$ by $\pfi$. We will control this component by fixing its gauge.

\section{Coisotropy form}\label{S2.5}

In this section we establish some elementary properties of the coisotropy form and compute it explicitly on $S^2$. Among other places it will feature prominently in the gauge theory of \refS{S3}.

We begin with a formal definition. Let $G$ be a Lie group equipped with a bi-invariant Riemannian metric and $H<G$ be a closed subgroup. Denote by $\h^\perp$ the orthogonal complement to the Lie algebra of $H$ in the induced invariant metric on $\g$. Let $x_0:=1H=\pi(1)\in G/H$ be the projection of the group identity, then the projection
$\g=T_1G\ovs{\pi_*}\lra T_{x_0}X$ identifies $\h^\bot$ with the
tangent space to $G/H$ at $x_0$. Left action of $G$ on $X:=G/H$ allows one
to extend the isomorphism of $\h^\bot$ to an arbitrary $T_xX$. Note that every vector in $T_xX$ has the form
$g(\pi_*\xi)$ for $\xi\in\g=T_1G$ (we take the liberty of writing $gT$ instead of $L_{g*}T$). 
\begin{definition}[Coisotropy form]\label{D:coisot}
The coisotropy form $\omega^\bot\in\Gamma(\Lambda^1X\otimes\g)$ of $X$ is
\bee\label{e0.41}
\omega^\bot
(g(\pi_*\xi)):=\Ad_*(g)\pr_{\h\bot}(\xi),
\eee
or equivalently
\bee\label{coisot}
\pi^*\omega^\bot:=\Ad_*(g)\pr_{\h\bot} (g^{-1}dg),
\eee
where $G\overset{\pi}{\lra}G/H$ is the quotient map and $\Ad_*(g)\eta:=g\eta g^{-1}$.
\end{definition}
\noindent Before proving elementary properties of the coisotropy form we recall some relevant algebraic notions for the convenience of the reader. The isotropy subgroup of a point $x\in X$ is 
$$
H_x:=\{\gamma\in G|\gamma x=x\}.
$$
If $x=gx_0=gH$ then $\gamma gH=gH$ is equivalent to $\gamma\in\Ad(g)H$ and
$$
H_x=\mbox{Ad}(g)H,\qquad x=gH.
$$ 
By analogy we define the isotropy subalgebra $\h_x$ of $x\in X$ and the coisotropy subspace $\h^\bot_x$:
\begin{equation*}
\bal
\h_x &:=\mbox{Ad}_*(g)\h,\quad x=gH\\
\h_x^\bot &:=\mbox{Ad}_*(g)\h^\bot 
\eal
\end{equation*}
These are well-defined since $\mbox{Ad}_*(gh)=\mbox{Ad}_*(g)\mbox{Ad}_*(h)$ and both $\h,\h^\bot$ are $\mbox{Ad}_*(H)$--invariant. More geometrically, let $\xi x$ denote the action of a vector in $\g$ on a point in $X$.
Since $G$ acts transitively, for each $x\in X$ the map $\xi\mapsto\xi x$ is
onto $T_xX$. Its kernel is exactly the isotropy subalgebra $\h_x$. The next lemma establishes some basic properties of the coisotropy form.
\begin{lemma}\label{L:2.1}
\noindent {\rm(i)} $ \omega^\bot$ is well-defined and $\omega^\bot (\xi x)=\pr_{\h_x^\bot}(\xi)$.

\noindent {\rm (ii)} $L^*_\gamma\omega^\bot=\mbox{Ad}_*(\gamma)\omega^\bot$, i.e. $\omega^\bot$ is
left--equivariant.

\noindent {\rm (iii)}$|\omega^\bot(S)|=|S|$ for any $S\in TX$.
\end{lemma}
\begin{proof}
{\rm (i)} Since $\xi x\in T_xX$ it has the form
$$
g(\pi_*\widetilde{\xi})=\xi x=\xi gH=g\widetilde{\xi}H,
$$
where $x=gH$. Thus, one can take
$\widetilde{\xi}=\Ad_*(g^{-1})\xi)$. Now by \eqref{e0.41}
\bee\label{e0.44}
\omega^\bot (\xi x)=\omega^\bot
(g(\pi_*\widetilde{\xi}))=\Ad_*(g)\pr_{\h^\bot}
(\widetilde{\xi})=\Ad_*(g)\pr_{\h^\bot}(\Ad_*(g^{-1})\xi)
\eee
By linear algebra, if $\m$ is a subspace of a Euclidean space and ${\rm U}$ is an
isometry then
\be
\pr_{{\rm U}\m}={\rm U}\pr_{\m}{\rm U}^*={\rm U}\pr_{\m}{\rm U}^{-1}.
\ee
Since $\Ad_*(g)$ is an isometry we obtain from \eqref{e0.44} that
\be
\omega^\bot(\xi x)=\pr_{\Ad_*(g)\h^\bot}(\xi)=\pr_{\h_x^\bot}(\xi).
\ee
Since the last expression depends only on $x\in X$
and not on $g\in G$ we conclude that $\omega^\bot$ is well defined.

{\rm (ii)} Since in our notation $L_{\gamma *}S=\gamma S$:
\begin{multline*}
L_\gamma^*\omega^\bot(g(\pi_*\xi))=\omega^\bot(\gamma
g(\pi_*\xi))=\Ad_*(\gamma g)\pr_{\h^\bot}(\xi)\\
=\Ad_*(\gamma)(\Ad_*(g)\pr_{\h^\bot}(\xi ))=\Ad_*(\gamma)\omega^\bot(g(\pi_*\xi)).
\end{multline*}

{\rm (iii)} Since $\Ad_*(\gamma)$ is an isometry and the metric on
$X$ is left-invariant it suffices to check the equality for $x=x_0$,
$g=1$. But there the lift of $S=\pi_*\xi$ is exactly
$\overline{S}=\pr_{\h^\bot}(\xi)$ since $\Ker \pi_*=\h$. Therefore
by definition of the Riemann quotient: $|S|:=|\overline{S}|=|\omega^\bot(S)|$.
\end{proof}
As one can see from \refL{2.1} the coisotropy form is just a way to rewrite tangent vectors on $X$ as vectors in $\g$ in an algebraically nice way. The Maurer-Cartan form $dg\,g^{-1}$ plays the same role on $G$. The next example gives an  explicit description for the case of $\CP^1=S^2=SU_2/U_1$.

\begin{example}[Coisotropy form of $\CP^1$]\label{E:CP1coisot}
Recall that $SU_2$ is represented by
\be
\begin{aligned}
SU_2&=\left\{\left.\begin{pmatrix} z &w\\
-\overline{w}&\overline{z}\end{pmatrix}\right|\ z, w\in\C,
|z|^2+|w^2|=1\right\}\\
U_1&=\left\{\left.\begin{pmatrix} z&0\\ 0&\overline{z}\end{pmatrix}\right|\ z\in\C,
|z|=1\right\}<SU_2
\end{aligned}
\ee
It is convenient to use the isomorphism $\begin{pmatrix}
z&w\\ -\overline{w}&\overline{z}\end{pmatrix}\longmapsto z+w
j\in\H$ with  the algebra of quaternions and use the quaternionic
notation. In this notation
\bee\label{quat}
\begin{aligned} 
G &=SU_2=\{q\in\H|\ |q|=1\}\\
H &=U_1=\{q\in\C|\ |q|=1\}\\
\g &=\su_2=\{\ q\in\H|\Re(q)=0\}=\Im\H\\
\h &=\uu_1=\{\ q\in\C |\Re(q)=0\}=\Im\C =i\R.
\end{aligned}
\eee
There is a useful embedding 
\begin{align*} 
\CP^1 &\ovs{\tau}\hra \H\\
qU_1 &\mapsto qiq^{-1}=\Ad_*(q)i
\end{align*} 
with the image
\be
\tau (\C P^1)=S^2=\{q\in\Im\H|\ |q|=1\}\subset\Im\H =\g.
\ee
It is convenient to identify $\CP^1$ with this image. 

We will now compute the coisotropy form under this identification. Since $\omega^\bot$ is
left-equivariant it suffices to compute it for $x_0=\pi(1)$ that is mapped into $i$ under $\tau$. 
Differentiating $\tau$ one gets 
\begin{align*} 
T_{x_0}\C P^1&\ovs{\tau_*}\lra T_iS^2\ ,\\
\xi x_0&\longmapsto [\xi ,i]
\end{align*} 
where as usual $T_iS^2$ is identified with a subspace in $\Im\H$. 
Therefore by \refL{2.1}(i)
$$
\omega_{x_0}^\bot (\xi
x_0)=\pr_{\h_{x_0}^\bot}(\xi)=\pr_{\h^\bot}(\xi)=\frac{1}{2}i[\xi
,i]=\frac{1}{2}i(\tau_*(\xi x_0)).
$$
Hence, if we identify $T_{x_0}\CP^1$ with $T_iS^2$ and write $\omega_i^\bot$
as a form on $\Im\H$ it becomes $\omega_i^\bot
(\eta)=\frac{1}{2}i\eta$. Analogously, identifying $T_x\CP^1$ with $T_{\tau(x)}S^2\subset\Im\H$ and using the left equivariance 
we get $\omega_x^\bot(\xi x)=\frac{1}{2}\tau (x)(\tau_*(\xi x))$. Thus,
\bee\label{e0.54}
\omega_q^\bot (\eta)=\frac{1}{2}q\eta,\quad q\in
S^2,\quad \eta\in T_qS^2.
\eee
Geometrically, this means that $\omega^\bot$ takes half of a vector in a tangent plane to $S^2$ and rotates it by $90^0$
counterclockwise in that plane. Its value is interpreted as an element of $\g=\Im\H=\R^3$.
\end{example}

\section{Gauge theory on coset bundles}\label{S3}

In this section we develop a systematic gauge interpretation of quantities that appear in the Faddeev-Skyrme models. Two principal bundles play special roles: the trivial one $M\times G$ and its subbundle, which is the pullback of the quotient bundle $G\to G/H$ by $\pfi$. Compared with general principal bundles such pullbacks, which we call {\it coset bundles}, admit many additional structures that they share with trivial bundles. After reviewing briefly trivial bundles (trivial connection, pure-gauge connections, global gauges, etc.) we proceed to coset bundles and develop a calculus for connections on them. To avoid technicalities we assume throughout this section that all quantities are $C^\infty$ smooth. Our notation is more or less standard, see \cite{MM}.

Trivial bundles are the simplest principal bundles \cite{Hus,MM} and their total spaces are 
products $P=M\times G$. The principal action is
multiplication by $G$ on the right in the second component 
$$
\bal
(M\times G)\times G &\lra M\times G\\
((m,g),\gamma)&\mapsto (m,g\gamma)
\eal
$$
and the projection is the projection $M\times G\ovs{\pi_1}\lra M$ to the first component. 
Trivial bundles and only those can be obtained by pullback from the bundle over one point $G\lra\pt$.
Indeed, in general pullback of a principal bundle $P\ovs{\pi}\lra X$ by a map $M\ovs{\pfi}\lra X$ is
\be
\pfi^* P:=\{(m,p)\in M\times P|\pfi (m)=\pi (p)\}
\ee
and for $P_{\pt}:=(G\to\pt)$ the defining condition trivializes leaving just $M\times
G$. 

For each pullback bundle there is a canonical bundle morphism
$M\times P \supset \pfi^* P\ovs{\pi_2}\lra P$ that allows to transfer connection forms: every connection $A$ on $P$ induces a connection $\pi_2^*A$ on $\pfi^*P$. For $P_{\pt}$ the
left-invariant Maurer-Cartan form $\theta_L=g^{-1}dg$ gives a
canonical connection and $\pi_2^*\theta_L$ (also denoted $g^{-1}dg$ when no confusion can result)
is called the {\it trivial connection} on $M\times G$. More
connections can be obtained by using gauge transformations (bundle automorphisms) $f$ of $M\times G$. 
Since $f(m,g\gamma)=(m,f_2(m,g)\gamma)$ we
have $f_2(m,g)=f_2(m,1)g=u(m)g$, where $M\ovs{u}\lra G$ and
$f(m,g)=(m,u(m)g)$. Conversely, any map into $G$ induces a gauge
transformation and we have a one-to-one correspondence between maps
$M\to G$ and $\Aut(M\times G)$. Applying them to the trivial
connection we get new ones:
\begin{multline}
f^*\pi_2^*(g^{-1}dg)=(\pi_2\circ f)^*(g^{-1}dg)=(ug)^{-1}d(ug)\\
=g^{-1}u^{-1}(dug+udg)=\Ad_*(g^{-1})(u^{-1}du)+g^{-1}dg.
\end{multline}
Such connections are called {\it pure-gauge} since they are trivial up to gauge equivalence (one could define
pure-gauge connections on any principal bundle relative to a reference
connection $A_0$ as those of the form $f^*A_0$ but this is not
common). Thus, we have a canonical choice of a reference connection
$A_0:=\pi_2^*(g^{-1}dg)=g^{-1}dg$ (by abuse of notation) and may consider differences $A-A_0$. 
The differences $A-A_0$ although horizontal are not
invariant under the right action of the structure group. We only have $\Ad_*$-equivariance:
$$
R_g^*(A-A_0)=\Ad_*(g^{-1})(A-A_0).
$$ 
On a trivial bundle (and, as we will see shortly, on a coset bundle) this can be fixed by a correction
factor $Ad_*(g)$. Indeed, the form $\Ad_*(g)(A-A_0)$ is horizontal, invariant
and therefore descends to a $\g$--valued form on $M$.
\begin{definition}[Gauge potentials on trivial bundles]\label{D:trivpot}
The gauge potential of a connection $A$ on $M\times G$ is
the form $a\in\Gamma(\Lambda^1\otimes\g)$ satisfying
\bee\label{trivpot}
\pi_1^*a=\Ad_*(g)(A-g^{-1}dg).
\eee
\end{definition}
It is immediate from \eqref{trivpot} that for pure-gauge connections
$A=f^*(g^{-1}dg)$ one gets $a=u^{-1}du$. Note that conventionally $a$ is introduced via local gauges and is also called connection in a local gauge \cite{MM,DFN} (of course, on trivial bundles local gauges are global). It is in this sense that $a$ is a pure-gauge connection in \cite{AK1,AK2}. We use the above construction because it conveniently generalizes to coset bundles while global gauges do not.

Curvature forms also descend to forms on the base. For the gauge potential 
$a$ of a connection $A$ define $F(a)$ by
\bee\label{e0.39}
\pi_1^*F(a)=\Ad_*(g)F(A).
\eee
Then a simple computation shows that
\bee\label{e0.40}
F(a)=da+a\wedge a.
\eee
Connections (potentials) with $F(A)=0$ $(F(a)=0)$ are
called {\it flat}. Every pure-gauge connection is flat as can be
seen directly from the expression $a=u^{-1}du$ for the potential.
The converse is true if $\pi_1(M)=0$, otherwise there is a topological
obstruction to constructing a {\it developing map} $u$ called holonomy \cite{AK1,KN}.

Now let us replace the one-point bundle $G\ovs{\i}\hra
G\ovs{\pi}\lra\pt$ by a quotient bundle $H\ovs{\i}\lra G\ovs{\pi}\lra
G/H=:X$. Most of the above generalizes to pullbacks of these bundles
under maps $M\ovs{\pfi}\lra X$.
\begin{definition}[Coset bundles]\label{D:coset}
A principal bundle is called a coset bundle if it is isomorphic to a pullback of a quotient bundle
$H\hra G\to G/H=X$, where $H<G$ is a closed subgroup of a Lie group $G$. Given $M\ovs{\pfi}\lra X$ 
we denote for short
$$
\pfi^* G:=\{(m,g)\in M\times G|\ \pfi(m)=gH\}\subset M\times G.
$$ 
\end{definition}
Any connection form $A$ on the trivial bundle $M\times G$ restricted to $\pfi^* G$ has the {\it
isotropy decomposition}:
\bee\label{e0.45}
A=\pr_\h A+\pr_{\h^\bot}A=:A^\vert+A^\bot.
\eee
Since $\Ad_*(h)$ commutes with $\pr_\h$ it follows from the definition of a principal connection that $A^\vert$ is a connection form on $\pfi^*G$. Therefore, the reference connection $A_0=g^{-1}dg$ on
$M\times G$ gives us a natural choice of a {\it reference connection on $\pfi^*G$}:
\bee\label{e0.46}
B_0:=A_0^\vert=(g^{-1}dg)^\vert=\pr_\h(g^{-1}dg)
\eee
Since $\pfi^*G\subset M\times G$ the correction
factor $\Ad_*(g)$ is still available and we can copy \refD{trivpot} to
set
\begin{definition}[Untwisted gauge potentials on coset bundles]\label{D:cospot}
The untwisted gauge potential of a connection $B$ on $\pfi^*G$ is
the form $b\in \Gamma (\Lambda^1M\otimes\g)$ satisfying
\bee\label{e0.47}
\Ad_*(g)(B-(g^{-1}dg)^\vert)=\pi_1^*b.
\eee
\end{definition}
Note that traditionally $B$ is represented by a gauge potential $\beta$, which is an $\Ad_*(\pfi^*G)$--valued $1$-form \cite{MM}. The latter bundle is usually non-trivial but has fiber $\h$. Our $b$ is a $\g$--valued form, i.e. a section of the trivial bundle but with a larger fiber $\g$. Thus, to untwist the potentials we pay the price of enlarging the fiber. Analogously, recall that traditionally gauge transformations are represented by sections of the bundle $\Ad(P)=P\times_{\Ad} H$ with fiber $H$. But on coset bundles they also can be untwisted into $G$--valued maps at the price of fiber extension. However, $b$ actually takes values in the isotropy subbundle $\h_\pfi$ and the gauge transformations are sections of $H_\pfi$, i.e. maps $M\ovs{w}\lra G$ that satisfy $w (m)\in H_{\pfi (m)}$, see \eqref{e0.10}.
\begin{lemma}\label{L:2.3}
There is an isometric isomorphism of vector bundles
$$
\begin{aligned}
\Ad_*(\pfi^*G)&\ovs{\sim}\lra\h_\pfi\\
(m,[g,\xi])&\longmapsto (m,\Ad_*(g)\xi)
\end{aligned}
$$ 
that induces isomorphisms on differential forms 
$$
\Gamma(\Lambda^kM\otimes\Ad_*(\pfi^*G))\simeq\Gamma
(\Lambda^kM\otimes\h_\pfi)\subset\Gamma (\Lambda^kM\otimes\g).
$$
The gauge potential $\beta$ of a connection $B$ is transformed 
by this isomorphism into its untwisted gauge potential $b$.
There is also an isomorphism 
$$
\bal 
\Ad(\pfi^*G)&\ovs{\sim}\lra H_\pfi\\
(m,[g,\lambda])&\longmapsto (m,\Ad (g)\lambda)
\eal
$$ 
that induces isomorphism of the gauge group $\Gamma (\Ad
(\pfi^*G))\ovs{\sim}\lra\Gamma (H_\pfi)$, i.e
\bee\label{e0.55}
\Gamma (H_\pfi)=\{M\ovs{w}\lra G|w \pfi =\pfi\}\simeq\Gamma (\Ad (\pfi^*G))
\eee
\end{lemma}
\begin{proof}
One can see that the map given by the first formula and the map
$$
\begin{aligned} 
\h_\pfi&\lra\Ad_*(\pfi^*G)\\
(m,\eta)&\longmapsto (m,[g,\Ad_*(g^{-1})\eta])
\end{aligned}
$$ 
are both well-defined and inverses of each other. Therefore, they are both isomorphisms, and they are isometric
because $\Ad_*(g)$ is an isometry. A straightforward calculation using Defintion \ref{D:cospot} shows that $\beta$ is transformed into $b$. The second isomorphism is proved as above with $\Ad$ in
place of $\Ad_*$. For the last claim note that $w(m)=ghg^{-1}$ for some $h\in H$ and $w (m)\pfi (m)=w
(m)gH=ghg^{-1}gH=gH=\pfi (m)$, the converse follows similarly.
\end{proof}
\noindent {\bf Notational convention}: {\it Since we have little use for the
traditional gauge potentials \cite{MM} from now on expressions 'gauge
potential' or 'potential' will refer to the untwisted ones of \refD{cospot}
unless otherwise stated. Since the isomorphism of \refL{2.3} is
isometric results stated in the literature for traditional potentials
(such as the Uhlenbeck compactness theorem that we use in
\refS{S4}) are trivially rephrased in terms of our untwisted ones. We utilize such rephrasings 
without special notice.}

Recall that we used the symbols $\vert,\bot$ in \refD{isotrop} to denote projections to $\h_\pfi$, $\h_\pfi^\bot$. Next Lemma shows that this is in line with our notation for connection forms. It also shows that the coisotropy form $\omega^\bot$ \eqref{coisot} appears naturally in the gauge theoretic context.
\begin{lemma}\label{L:2.4} Let $a$ be the untwisted gauge potential of a connection $A$ on a coset bundle. Then $a^\vert, a^\bot$ are characterized by
\bee\label{e0.48}
\begin{aligned}
\pi_1^*a^\vert&:=\Ad_*(g)(A^\vert-(g^{-1}dg)^\vert),\\
\pi_1^*a^\bot &:=\Ad_*(g)(A^\bot -(g^{-1}dg)^\bot).
\end{aligned}
\eee
Moreover, if $A_u$ is a pure-gauge connection with potential $a_u=u^{-1}du$ then
\bee\label{mapcon}
 a_u^\bot=\Ad_*(u^{-1})(u\pfi)^*\omega^\bot -\pfi^*\omega^\bot.
\eee
\end{lemma}
\begin{proof}
Let $\pfi (m)=gH$ then $\pr_{\h_{\pfi
(m)}}=\pr_{\Ad_*(g)\h}=\Ad_*(g)\pr_\h\Ad_*(g^{-1})$ and since
$(m,g)\in\pfi^*G$ always satisfies $\pfi (m)=gH$ we have
\be
\begin{aligned}
\pi_1^*a^\vert=\pi_1^*(\pr_{\h_\pfi}(a))&=\pr_{\Ad_*(g)\h}(\pi_1^*a)=\Ad_*(g)\pr_\h\Ad_*(g^{-1})(\Ad_*(g)(A-g^{-1}dg))\\
&=\Ad_*(g)\pr_\h (A-g^{-1}dg)=\Ad_*(g)(A^\vert-(g^{-1}dg)^\vert).
\end{aligned}
\ee
The second formula follows from $a^\bot =a-a^\vert$. Since $\pi_1^*$ is mono these equalities characterize the components. For the third formula note that $A_u=f_2^*(g^{-1}dg)$, where
$f_2$ is the second component of the gauge transformation
$$
\begin{aligned} 
M\times G&\ovs{f} \lra M\times G\\
(m,g)&\longmapsto (m,u(m)g)
\end{aligned}
$$
It is easy to see by inspection that the following diagram commutes:
$$
\begin{diagram}
\pfi^* G & \rTo{f_2}  &G       \\
 \dTo{\pi_1}   &         &\dTo{\pi} \\
M           & \rTo{u\pfi}     & X
\end{diagram}
$$
Therefore, 
\bee\label{e0.50}
\begin{aligned}
\pi_1^*(\Ad_*(u^{-1})&(u\pfi)^*\omega^\bot)=\Ad_*((u\circ\pi_1)^{-1})\pi_1^*(u\pfi)^*\omega^\bot &&\\
&=\Ad_*((u\circ\pi_1)^{-1})f_2^*\pi^*\omega^\bot &&\\
&=\Ad_*((u\circ\pi_1)^{-1})f_2^*\Ad_*(g)(g^{-1}dg)^\bot && \text{by \eqref{coisot}}\\
&=\Ad_*((u\circ\pi_1)^{-1})\Ad_*((u\circ\pi_1)g)(f_2^*(g^{-1}dg)^\bot) && \text{since $f_2={(u\circ\pi_1)g}$}\\
&=\Ad_*(g)A_u^\bot &&
\end{aligned}
\eee
When $u$ is the constant $1$ map this equality turns into
\bee\label{e0.51}
\pi_1^*(\pfi^*\omega^\bot)=\Ad_*(g)(g^{-1}dg)^\bot
\eee
Subtracting \eqref{e0.51} from \eqref{e0.50} and using \eqref{e0.48} we get
the desired equality.
\end{proof}
\begin{example}[Isotropy decomposition on $\CP^1$]\label{E:CP1isotrop}
Recall from \refE{CP1coisot} that on $\CP^1=SU_2/U_1$ we can identify $\su_2$ with the space $\Im\H$ of
purely imaginary quaternions and $\uu_1$ with $i\R\subset\Im\H$. Therefore
\be
\begin{aligned}
\pr_\h (\xi)&=(\xi ,i)i=\Re (\xi
\overline{i})i=\frac{\xi\overline{i}+\overline{\xi}i}{2}i=-\frac{\xi
i+i\xi}{2}i=\frac{1}{2}(\xi -i\xi i)\\
\pr_{\h^\bot}(\xi)&=\xi -\frac{1}{2}(\xi -i\xi i)=\frac{1}{2}(\xi
+i\xi i)=\frac{1}{2}i(-i\xi +\xi i)=\frac{1}{2}i[\xi ,i],
\end{aligned}
\ee
where $\h^\bot =\uu_1^\bot$ is the linear span of\ $j$, $k$. Also recall that 
we can identify $\CP^1$ itself with the unit sphere $S^2$ in $\Im\H$. Under this
identification a map $M\ovs{\pfi}\lra\CP^1$ turns into a map $M\ovs{\phi}\lra S^2$ with
\be
\phi(m):=qiq^{-1}=qi\overline{q},\quad \text{if $\pfi (m)=qU_1$.}
\ee
With this notation:
\be
\pr_{\h_\pfi}(\xi)=\Ad_*(q)\pr_\h(\Ad_*(q^{-1})\xi)=\Ad_*(q)(\Ad_*(q^{-1})\xi,i)i
=(\xi ,\Ad_*(q)i)\Ad_*(q)i=(\xi,\phi)\phi
\ee
since $\Ad_*(q)$ is an isometry and $(\xi ,\eta)\in\R$ and
therefore commutes with all quaternions. Analogously,
$$
\pr_{\h_\phi^\bot}(\xi)=\frac{1}{2}\phi[\xi,\phi].
$$ 
Thus, by \eqref{e0.49} we get in terms of $\phi$:
\bee\label{e0.53}
a^\vert=(a,\phi)\phi,\quad a^\bot =\frac{1}{2}\phi
[a,\phi]. \eee
These are the expressions used in \cite{AK2}.
\end{example}
Note that the last claim of \refL{2.3} gives a gauge description of $\pfi$-stabilizing maps. Therefore, it is natural to investigate their properties further. The main role of gauge transformations is their action on connection forms -- the gauge action. As connections are now represented by (untwisted) gauge potentials $b\in\Gamma (\Lambda^1M\otimes\g)$ (\refD{cospot}) and gauge transformations by maps $M\ovs{w}\lra G$, we
would like to have an explicit expression for the action of $w$
on $b$. Similarly, curvature of a connection $B$ on $\pfi^*G$ is a
horizontal equivariant $2$-form on $\pfi^*G$ and after applying the
correction factor $\Ad_*(g)$ we can make it invariant and descend it to
$M$. Again, we would like an explicit expression for the result in terms of
the potential $b$. This prompts the following definition.
\begin{definition}[{\bf Gauge action and curvature for gauge potentials}]\label{D:bwF}
Let $f_w$ be the gauge transformation corresponding to the map
$M\ovs{w}\lra G$, $w\in\Gamma (H_\pfi)$ and $b$ be the potential of a connection $B$. Then $b^w$
denotes the gauge potential of the transformed connection $f_w^*B$. The curvature potential $F(b)$ is defined by
\bee\label{e0.56}
\pi_1^*F(b)=\Ad_*(g)F(B)=\Ad_*(g)(dB+B\wedge B).
\eee
\end{definition}
Obviously, $F(b)\in\Gamma(\Lambda^2 M\otimes\g)$, moreover $F(b)\in\Gamma
(\Lambda^2M\otimes\h_\pfi)$ since $dB+B\wedge B$ is $\h$--valued.
Note that usually $F(\beta)$ is defined for a twisted potential $\beta$
and is an $\Ad_*(P)$--valued $2$-form descended from $F(B)$. This $F(\beta)$ corresponds to our $F(b)$ 
under the induced isomorphism of \refL{2.3}.

Before we derive explicit expressions for $b^w$, $F(b)$ let us make several preparations. 
First, it is convenient to extend the notation $\vert$, $\bot$ to all $\g$--valued forms on
$\pfi^*G$ and $M$:
\bee\label{e0.57}
\bal 
R^\vert &:=\pr_\h (R)  && \text{for $R\in\Gamma(\Lambda^\bullet(\pfi^*G)\otimes\g)$}\\
R^\bot &:=\pr_{\h^\bot}(R) &&\\
r^\vert &:=\pr_{\h_\pfi}(r)  && \text{for $r\in\Gamma(\Lambda^\bullet M\otimes\g)$.}\\
r^{\bot} &:=\pr_{\h_\pfi^\bot}(r) && 
\eal
\eee
By \eqref{e0.49}, \eqref{e0.45} this agrees with our previous notation for $A$ and $a$.

Second, note that every connection form $B$ on $\pfi^*G$ is the isotropic part of a
(non-unique) connection $A$ on $M\times G$. This is easy
to see using the gauge potentials $b$. By \refD{cospot} one has that $b$ is an $\h_\pfi$--valued $1$--form, 
but $\h_\pfi\subset\g$ and it can also be treated as a $\g$--valued one. By \refD{trivpot} any 
$\g$--valued $1$--form represents a connection on $M\times G$. Let $A$ denote this connection for 
$b$ treated as a $\g$--valued form then $B=A^\vert$ as required. More explicitly we have
\be
\bal
\pi_1^*b &=\Ad_*(g)(B-(g^{-1}dg)^\vert) && \text{on $\pfi^*G\subset M\times G$}\\ 
\pi_1^*a &=\Ad_*(g)(A-g^{-1}dg)  && \text{on  $M\times G$},
\eal
\ee
and therefore 
\bee\label{minext}
A=B+(g^{-1}dg)^\bot
\eee
on $\pfi^*G$. It can be uniquely extended to the entire $M\times G$ by equivariance.
This is the {\it minimal extension} of $B$. More generaly, we could take any $\h_\pfi^\bot$--valued $1$-form $\delta$ on $M$, set $a=b+\delta$ and take $A$ on $M\times G$ that corresponds to $a$.

Third, the gauge transformation $f_w$ from \refD{bwF} can be found explicitly. By \refL{2.3} $w$
corresponds to a section $\sigma$ of $\Ad(\pfi^*G)$ given by
$$
\sigma(m):=(m,[g,\Ad_*(g^{-1})w (m)])
$$
In its turn, by the isomorphism between $\Gamma (\Ad (\pfi^*G))$ and $\Aut(\pfi^*G)$ this section corresponds to
$$
f_w(m,g)=(m,g\Ad_*(g^{-1})w(m))=(m,w(m)g).
$$ 
Although we obtained it as a gauge transformation of $\pfi^*G$ only, 
it obviously extends to a gauge transformation of $M\times G$ that we denote
by the same symbol. If $A$ is a connection on $M\times G$ with the gauge potential $a$ then 
the gauge potential $a^w$ of $f_w^*A$ is easily found to be \cite{DFN,MM}:
\bee\label{e0.59}
a^w =\Ad_*(w^{-1})a+w^{-1}dw .
\eee
Now we are ready to derive the promised formulas. The idea of the proof is to extend a connection on $\pfi^*G$ to a connection on $M\times G$, use the well-known formulas for potentials on a trivial bundle and then project them to the potentials on a coset bundle. The coisotropy form $\omega^\bot$ makes an important appearence here.
\begin{theorem}\label{T:2.1}
Let $B$ be a connection on $\pfi^*G$, $b$ be its (untwisted) gauge potential and
$w$ be a section of $H_\pfi\subset M\times G$. Then
\bee\label{e0.60}
\bal
&{\rm (i)}\ \ \ \ \ \ b^w=\Ad_*(w^{-1})b+w^{-1}dw-(\Ad_*(w^{-1})-I)\pfi^*\omega^\bot\\
&{\rm (ii)}\ F(b^w)=\Ad_*(w^{-1})F(b)\\
&{\rm (iii)}\ \ F(b)=db+b\wedge b-[b,\pfi^*\omega^\bot]-(\pfi^*\omega^\bot\wedge\pfi^*\omega^\bot)^\vert.
\eal
\eee
\end{theorem}
\begin{proof}
{\rm (i)} Let $A$ be the minimal extension of $B$ to $M\times G$ then we have for the potentials $a,b$
then
\begin{align*}
\pi_1^*a^w &=\Ad_*(g)(f_w^*A-g^{-1}dg) && \text{by \refD{trivpot}}\\
\intertext{and}
\pi_1^*b^w &=\pi_1^*(a^\vert)^w=\Ad_*(g)(f_w^*A^\vert-(g^{-1}gd)^\vert) && \text{by \refD{cospot}}\\
&=\Ad_*(g)((f_w^*A)^\vert-(g^{-1}dg)^\vert) && \text{since $\pr_\h$ commutes with $f_w^*$}\\
&=\pi_1^*(a^w)^\vert && \text{by \refD{isotrop}.}
\end{align*}
Therefore $b^w=(a^w)^\vert$. Since $\pr_{\h_\pfi}$ commutes with $\Ad_*(w^{-1})$ 
for $w\in\Gamma (H_\pfi)$ we have further
$$
b^w=(a^w)^\vert=(\Ad_*(w^{-1})a+w^{-1}dw)^\vert=\Ad_*(w^{-1})a^\vert+(w^{-1}dw)^\vert.
$$
But by definition of the minimal extension $b=a^\vert=a$ and 
\bee\label{e0.63}
b^w=\Ad_*(w^{-1})b+(w^{-1}dw)^\vert.
\eee
When $w\pfi =\pfi$ the equality \eqref{mapcon} becomes
\bee\label{e0.62}
(w^{-1}dw)^\bot =\Ad_*(w^{-1})\pfi^*\omega^\bot-\pfi^*\omega^\bot
\eee
and therefore
$$
(w^{-1}dw)^\vert=w^{-1}dw-(w^{-1}dw)^\bot=w^{-1}dw-(\Ad_*(w^{-1})-I)\pfi^*\omega^\bot
$$
Substituting this into \eqref{e0.63} we get the required formula.

{\rm (ii)} For any horizontal equivariant form $R$ on $\pfi^*G$ one has 
$\Ad_*(g)R=\pi_1^*r$ with a unique form $r$ on $M$. We claim that then
\bee\label{e0.61}
\Ad_*(g)(f_w^*R)=\pi_1^*(\Ad_*(w^{-1})r).
\eee
Indeed,
$$
f_w^*(\Ad_*(g)R)=\Ad_*((w\circ\pi_1)g)f_w^*R=\Ad_*(w\circ\pi_1)(\Ad_*(g)f_w^*R)
$$
and
\begin{multline*}
\Ad_*(g)(f_w^*R)=\Ad_*((w\circ\pi_1)^{-1})f_w^*(\pi_1^*r)=\Ad_*((w\circ\pi_1)^{-1})(\pi_1\circ f_w)^*r\\
=\Ad_*((w\circ\pi_1)^{-1})\pi_1^*r=\pi_1^*(\Ad_*(w^{-1})r).
\end{multline*}
Applying \eqref{e0.61} to $R=F(B)=dB+B\wedge B$ one obtains
\be
\Ad_*(g)F(f_w^*B)=\Ad_*(g)(f_w^*F(B))=\pi_1^*(\Ad_*(w^{-1})F(b))=\pi_1^*F(b^w),
\ee
which implies {\rm (ii)} since $\pi_1^*$ is mono. 

{\rm (iii)} Again, let $A$ be the minimal extension of $B$. For potentials $a=b$ 
we now have two different curvatures: one induced from the curvature of $A$ by \eqref{e0.39}, 
the other induced from the curvature of $B$ by \eqref{e0.56}. To avoid confusion we denote the former $\widehat{F}(a)$ for the duration of this proof only. Thus,
\begin{align*}
\pi_1^*F(b)=\pi_1^*F(a) &=\Ad_*(g)(dB+B\wedge B)\\
\pi_1^*\widehat{F}(a) &=\Ad_*(g)(dA+A\wedge A).
\end{align*}
Since $A$ is the minimal extension by \eqref{minext}
\be
\bal
dA &=dB+d(g^{-1}dg)^\bot\\
A\wedge A &=B\wedge B+[B,(g^{-1}dg)^{\bot}]+(g^{-1}dg)^\bot\wedge (g^{-1}dg)^\bot
\eal
\ee
Since $g^{-1}dg$ is flat it satisfies
$$
d(g^{-1}dg)=-(g^{-1}dg)\wedge(g^{-1}dg)
$$
Decomposing $g^{-1}dg=(g^{-1}dg)^\vert+(g^{-1}dg)^\bot$ and taking into account \eqref{shhfi} we get
\be
d(g^{-1}dg)^\bot=-(g^{-1}dg\wedge g^{-1}dg)^\bot
=-[(g^{-1}dg)^\vert,(g^{-1}dg)^\bot]-((g^{-1}dg)^\bot\wedge (g^{-1}dg)^\bot)^\bot.
\ee
Putting it together:
\begin{multline*}
dA+A\wedge A=dB+d(g^{-1}dg)^\bot+B\wedge B+[B,(g^{-1}dg)^{\bot}]+(g^{-1}dg)^\bot\wedge (g^{-1}dg)^\bot\\
=dB+B\wedge B+[B,(g^{-1}dg)^\bot]+(g^{-1}dg)^\bot\wedge (g^{-1}dg)^{\bot}\\
-[(g^{-1}dg)^\vert,(g^{-1}dg)^\bot]-((g^{-1}dg)^\bot\wedge
(g^{-1}dg)^\bot)^\bot\\
=dB+B\wedge B+[(B-(g^{-1}dg)^\vert),(g^{-1}dg)^\bot]+((g^{-1}dg)^\bot\wedge (g^{-1}dg)^\bot)^\vert.
\end{multline*}
Now apply $\Ad_*(g)$ to both sides and distribute it under $\wedge$ and $[\cdot ,\cdot]$ operations. Then we can interchange $\Ad_*(g)$ with the $\vert,\perp$ signs using that $\Ad_*(g)\pr_\h =\pr_{\h_\pfi}\Ad_*(g)$. Since 
$$
\Ad_*(g)(B-(g^{-1}dg)^\vert)=\pi_1^*b
$$
by \eqref{e0.48} and
$$
\Ad_*(g)(g^{-1}dg)^\bot =\pi_1^*(\pfi^*\omega^\bot)
$$ 
by \eqref{e0.51} the equality turns into
\be
\pi_1^*\widehat{F}(a)=\pi_1^*F(b)+\pi_1^*[b,\pfi^*\omega^\bot]+\pi_1^*(\pfi^*\omega^\bot\wedge\pfi^*\omega^\bot)^\vert
\ee
Removing $\pi_1^*$ and recalling that $\widehat{F}(a)=da+a\wedge a=db+b\wedge b$ by \eqref{e0.40} we get the required formula.
\end{proof}
\begin{remark*}
Note that the formulas from \refT{2.1} look like their analogs for trivial bundles with correction terms depending on the pullback of the coisotropy form $\pfi^*\omega^\bot$. If $\pfi$ is a constant map and the bundle $\pfi^*G$ is trivial then $\pfi^*\omega^\bot=0$ and we recover the formulas for trivial bundles.
\end{remark*}
An interesting consequence of \refT{2.1} is
\begin{corollary}\label{C:Dafi}
\bee\label{Dafi} 
(a^w)^\bot +\pfi^*\omega^\bot
=\Ad_*(w^{-1})(a^\bot +\pfi^*\omega^\bot) 
\eee
\end{corollary}
\begin{proof}
By direct computation from \eqref{e0.63}
\be
\bal
(a^w)^\bot 
&=a^w-(a^w)^\vert=\Ad_*(w^{-1})a +w^{-1}dw-\Ad_*(w^{-1})a^\vert -(w^{-1}dw)^\vert\\
&=\Ad_*(w^{-1})a^\bot +(w^{-1}dw)^\bot\\ 
&=\Ad_*(w^{-1})a^\bot+\Ad_*(w^{-1})\pfi^*\omega^\bot-\pfi^*\omega^\bot\\ 
&=\Ad_*(w^{-1})(a^\bot+\pfi^*\omega^\bot)-\pfi^*\omega^\bot.
\eal
\ee
\end{proof}
Comparing \eqref{Dafi} to \eqref{e0.60}(ii) we see that the quantity $a^\bot
+\pfi^*\omega^\bot$ transforms like curvature. This reflects the following situation for connections.
In a principal bundle the only local gauge-equivariant functional of a connection $A$ is its curvature $F(A)$.
Equivariance refers to the gauge action induced by that {\it same bundle}. On the other hand, 
if we consider the gauge action induced by a {\it subbundle}
the curvature is joined by the coisotropic part $A^\bot$ with respect to this subbundle. It follows from \eqref{e0.48} and \eqref{e0.51} that
\bee\label{e0.65}
\Ad_*(g)A^\bot =\pi_1^*(a^\bot
+\pfi^*\omega^\bot).
\eee
Such partial gauge equivalence arises in {\it nonlinear $\sigma$-models} of quantum physics \cite{BMSS}. 
The gauge principle implies in this situation that physical Lagrangians should be functions of $a^\bot +\pfi^*\omega^\bot$ and $F(a^\vert)$. Faddeev-Skyrme functionals rewritten for potentials depend on the first quantity only, see \eqref{e0.13}.

Projecting \eqref{e0.60} to $\h_\pfi$, $\h_\pfi^\bot$ and taking into account \eqref{shhfi} we get
\begin{corollary} 
For any gauge potential on a coset bundle $\pfi^*G$ one has
\bee\label{e0.67} 
\bal 
F(b) &=(db)^\vert+b\wedge b-(\pfi^*\omega^\bot\wedge\pfi^*\omega^\bot)^\vert\\
(db)^\bot &=[\pfi^*\omega^\bot ,b]
\eal 
\eee
\end{corollary}
Taking $\pfi=\id_X$ and $b=0$ in \eqref{e0.60}(iii) corresponds to computing the curvature potential of the
reference connection $(g^{-1}dg)^\vert$ on the quotient bundle $H\hra G\to G/H=X$.
\begin{corollary}\label{C:refcurv}
The curvature potential of the reference connection $(g^{-1}dg)^\vert$
on $G\ovs{\pi}\lra X$ is
\bee\label{e0.66}
F(0)=-(\omega^\bot\wedge\omega^\bot)^\vert.
\eee
\end{corollary}
This is another indication of a role that the coisotropy form plays
in geometry of homogeneous spaces. It becomes especially nice for 
symmetric spaces. Recall that $G/H$ is a {\it Riemannian symmetric space} if there is a homomorphic involution $G\to G$ that fixes $H$ pointwise \cite{Ar,Hl}. What is important to us is that in addition to the usual relations \eqref{shh} in a symmetric space one also has
\bee\label{snh}
[\h^\bot,\h^\bot]\subset\h,
\eee
and therefore
\bee\label{snhfi}
[\h_\pfi^\bot,\h_\pfi^\bot]\subset\h_\pfi.
\eee
Thus, \eqref{e0.66} becomes 
$$
F(0)=-\omega^\bot\wedge\omega^\bot.
$$

So far we derived formulas for all gauge potentials on $M\times G$. But, as follows from a direct computation, the pure-gauge potentials $a=u^{-1}du$ are in addition {\it flat}, i.e. $F(a)=da+a\wedge a=0$. The next Lemma translates this relation into equalities satisfied by $a^{\|}$ and $a^\bot$. They will be used in \refS{S5} to obtain apriori Sobolev estimates on $F(a^{\|})$ and $da^\bot$ in terms of the Faddeev-Skyrme functional.

It will be convenient to denote $\Phi:=\pr_{\h_\pfi}$ and treat it as an $\mbox{End}(\g)$--valued
function with $d\Phi\in\Gamma(\Lambda^1M\otimes\mbox{End}(\g))$. Differentiating the obvious by \eqref{shhfi} relation $\Phi a^{\|}=a^{\|}$ we get
\bee\label{e0.100}
d\Phi\wedge a^{\|}=(I-\Phi)da^{\|}=(da^{\|})^\bot.
\eee
Analogously differentiating $(I-\Phi)a^\bot=a^\bot$ yields
\bee\label{e0.101}
d\Phi\wedge a^\bot =-\Phi(da^\bot)=-(da^\bot)^{\|}.
\eee
When $X=G/H$ is a symmetric space one can do better. By \eqref{shhfi}
$I-\Phi=\pr_{\h^\bot_\pfi}$ and we have immediately
\bee\label{e0.109}
(I-\Phi)(a^\bot\wedge a^\bot)=0.
\eee
Differentiating \eqref{e0.109} gives a second relation
\bee\label{e0.110}
(I-\Phi)d(a^\bot\wedge a^\bot)=d\Phi\wedge(a^\bot\wedge a^\bot).
\eee
\begin{lemma}\label{L:dcurv}
Let $a$ be a flat gauge potential on $M\times G$, i.e. $da+a\wedge a=0$. Then
\bee\label{dcurv}
\bal
&{\rm (i)}\ F(a^{\|})=d\Phi\wedge
a^\bot-\Phi(a^\bot\wedge
a^\bot)-\Phi(\pfi^*\omega^\bot\wedge\pfi^*\omega^\bot)\\
&{\rm (ii)}\ da^\bot=-d\Phi\wedge a^{\|}-d\Phi\wedge
a^\bot-[a^{\|},a^\bot]-(I-\Phi)(a^\bot\wedge a^\bot).
\eal
\eee
If moreover $X=G/H$ is a Riemannian symmetric space then
\bee\label{sdcurv}
\bal
&{\rm (i')}\ F(a^{\|})=d\pfi\wedge a^\bot-a^\bot\wedge a^\bot-\pfi^*\omega^\bot\wedge\pfi^*\omega^\bot\\
&{\rm (ii')}\ da^\bot=-d\Phi\wedge a^{\|}-d\Phi\wedge a^\bot-[a^{\|},a^\bot]\\
&{\rm (iii')}\ d(a^\bot\wedge a^\bot)=-[d\Phi\wedge a^{\|},a^\bot]+d\Phi\wedge(a^\bot\wedge a^\bot).
\eal
\eee
\end{lemma}
\begin{proof}
{\rm (i)} By the product rule and flatness:
\be
\bal da^{\|}&=d(\pfi a)=d\Phi\wedge a+\pfi (da)=d\Phi\wedge a-\Phi(a\wedge a)\\
&=d\Phi\wedge a-\Phi((a^{\|}+a^\bot)\wedge(a^{\|}+a^\bot))\\
&=d\Phi\wedge a-\Phi(a^{\|}\wedge a^{\|}+[a^{\|},a^\bot]+a^\bot\wedge a^\bot).
\eal
\ee 
Since $\alpha\wedge\alpha=1/2[\alpha,\alpha]$ by \eqref{shhfi} the form $a^{\|}\wedge a^{\|}$ takes values in $\h_\pfi$ and
$[a^{\|},a^\bot]$ in $\h_\pfi^\bot$. Therefore 
$$
\Phi(a^{\|}\wedge a^{\|})=a^{\|}\wedge a^{\|}\quad \text{and} \quad \Phi[a^{\|},a^\bot]=0.
$$
Thus we get 
\bee\label{e0.103}
da^{\|}+a^{\|}\wedge a^{\|}=d\Phi\wedge
a^{\|}+d\Phi\wedge a^\bot-\Phi(a^\bot\wedge a^\bot).
\eee
By \eqref{e0.67}:
\be
\bal
F(a^{\|}) &=(da^{\|})^{\|}+a^{\|}\wedge a^{\|}-(\pfi^*\omega^\bot\wedge\pfi^*\omega^\bot)^{\|}\\
          &=\Phi(da^{\|}+a^{\|}\wedge a^{\|}-\pfi^*\omega^\bot\wedge\pfi^*\omega^\bot).
\eal
\ee
Subtracting $\pfi^*\omega^\bot\wedge\pfi^*\omega^\bot$ from both
sides of \eqref{e0.103}, applying $\Phi$ and taking into account that
$\Phi(d\Phi\wedge a ^{\|})=0$ by \eqref{e0.100} we get {\rm (i)}.

\noindent {\rm (ii)} Plugging $a=a^{\|}+a^\bot$ into $da+a\wedge a=0$ one gets
\be
da^\bot+a^\bot\wedge a^\bot+da^{\|}+a^{\|}\wedge a^{\|}+[a^{\|},a^\bot]=0.
\ee
Now rewriting $da^{\|}+a^{\|}\wedge
a^{\|}$ by \eqref{e0.103} and taking all terms except $da^\bot$ to the
righthand side gives {\rm (ii)}.

\noindent ${\rm (i')}$, ${\rm (ii')}$ follow directly from {\rm (i)}, {\rm (ii)} above and
\eqref{e0.109}.

\noindent ${\rm (iii')}$ Note that for odd degree forms $d(\alpha\wedge\alpha)=[d\alpha ,\alpha]$. 
Therefore, from {\rm (ii)}
\begin{multline}\label{daa}
d(a^\bot\wedge a^\bot)=[da^\bot,a^\bot]\\
=-[d\Phi\wedge a^{\|},a^\bot]-[d\Phi\wedge a^\bot,a^\bot]-[[a^{\|},a^\bot],a^\bot]
-[(I-\Phi)(a^\bot\wedge a^\bot),a^\bot].
\end{multline}
Since $d\Phi\wedge a^\bot=-\Phi(da^\bot)$ takes values in $\h_\pfi$ and
$[a^{\|},a^\bot]$ in $[\h_\pfi,\h_\pfi^\bot]\subset\h_\pfi$ we have that
$$
[d\Phi\wedge a^\bot,a^\bot]+[[a^{\|},a^\bot],a^\bot]
$$
is $\h_\pfi^\bot$--valued. On the other hand,
$$
d\Phi\wedge a^{\|}=(I-\Phi)da^{\|}
$$ 
is $\h_\pfi^\bot$--valued and by \eqref{snhfi} $[d\Phi\wedge a^{\|},a^\bot]$ takes values in $\h_\pfi$. Thus,
\begin{align*}
\Phi d(a^\bot\wedge a^\bot) &=-[d\Phi\wedge a^{\|},a^\bot]\\ 
(I-\Phi)d(a^\bot\wedge a^\bot) &=-[d\Phi\wedge a^\bot,a^\bot]-[[a^{\|},a^\bot],a^\bot]. 
\end{align*} 
Adding them together and using \eqref{e0.109} gives ${\rm (iii')}$.
\end{proof}

\section{Sobolev spaces and homotopy sectors}\label{S4}

In this section we give precise definitions of Sobolev spaces that we work with and establish some of their properties. Based on them we show that the description of homotopy classes given in \refS{S2} generalizes to our Sobolev maps and the variational problem for them makes sense.

Note that the Faddeev-Skyrme density 
\bee\label{e0.90}
e(\psi):=\frac{1}{2}|\psi^*\omega^\bot|^2+\frac{1}{4}|\psi^*\omega^\bot\wedge\psi^*\omega^\bot|^2
\eee
is defined almost everywhere for any $\psi\in W^{1,2}(M,X)$. Of
course it does not have to be integrable and we define the space of {\it finite energy maps}:
\bee\label{e0.91}
\bal
W_E^{1,2}(M,X):&=\{\psi\in
W^{1,2}(M,X)|e(\psi)\in L^1(M,\R)\}\\
&=\{\psi\in W^{1,2}(M,X)|E(\psi)<\infty\}.
\eal
\eee
Neither $W^{1,2}(M,X)$ nor $W_E^{1,2}(M,X)$ are Banach spaces or
even convex subsets of a Banach space. The word space here only means a topological space.

Since $\pi_2(G)=0$ smooth maps are dense in $W^{1,2}(M,G)$ but
not in $W^{1,2}(M,X)$ because $\pi_2(X)\neq0$ \cite{HL2}. This means in particular that formulas
derived for smooth maps can not be extended to Sobolev maps into $X$ simply
by smooth approximation. For instance, we can extend formula
\eqref{mapcon} to $u\in W^{1,2}(M,G)$, but we have to keep $\pfi$ smooth, or at least $C^1$.

We now want to define homotopy classes for $W_E^{1,2}(M,X)$ maps that we call homotopy sectors to avoid confusion. Motivated by \refT{2.1} we set 
\begin{definition}[$2$--homotopy sector]
Two maps $\pfi ,\psi\in W_E^{1,2}(M,X)$ are in the same $2$--homotopy sector
if there is a map $u\in W^{1,2}(M,G)$ such that $\psi=u\pfi$ a.e.
\end{definition}
Note that if $N$ is compact then $W^{1,2}(M,N)\subset L^\infty(M,N)$.
Therefore the product rule and the Sobolev multiplication theorems
\cite{Pl} imply that $W^{1,2}(M,G)$ is a group that acts on
$W^{1,2}(M,X)$. In particular, $W_E^{1,2}(M,X)$ is divided into
disjoint $2$--homotopy sectors. However, $W^{1,2}(M,G)$ no longer
acts on $W_E^{1,2}(M,X)$. In fact, even if $\pfi$ is smooth and $u\in
W^{1,2}(M,G)$ the product $\psi=u\pfi$ may not have finite Faddeev-Skyrme
energy. Indeed, by \eqref{mapcon}
\bee\label{e0.92}
\bal
\psi^*\omega^\bot&=\Ad_*(u)((u^{-1}du)^\bot+\pfi^*\omega^\bot)\\
\psi^*\omega^\bot\wedge\psi^*\omega^\bot&=\Ad_*(u)((u^{-1}du)^\bot\wedge(u^{-1}du)^\bot+[(u^{-1}du)^\bot
,\pfi^*\omega^\bot]+\pfi^*\omega^\bot\wedge\pfi^*\omega^\bot).
\eal
\eee
Therefore, $E(\psi)<\infty$ is equivalent to
$$
(u^{-1}du)^\bot\wedge(u^{-1}du)^\bot\in L^2(\Lambda^2M\otimes\g),
$$
which does not hold for arbitrary $u\in W^{1,2}(M,G)$. 

We shall see that the space $W_E^{1,2}(M,X)$ is to large from the topological perspective so we restrict it further to $\mathcal{E}(M,X)$. This space is defined in terms of $u$ or rather $a=u^{-1}du$ since all maps we have to consider in the process of minimization are of the form $u\pfi$. It follows from the results of \cite{AK3} that  $W_E^{1,2}(M,S^2)=\mathcal{E}(M,S^2)$, but if this is true in general is an open problem.
\begin{definition}[admissible maps]\label{D:admis}
A gauge potential $a$ is admissible if
\bee\label{admis}
\bal
&{\rm
1)}\ a^\bot\in
L^2(\Lambda^1M\otimes\g),\\
&{\rm 2)}\ a^\bot\wedge a^\bot\in L^2(\Lambda^2M\otimes\g),\\
&{\rm 3)}\ a^{\|}\in W^{1,2}(\Lambda^1M\otimes\g).
\eal
\eee
The space of admissible potentials is denoted
$\mathcal{E}(\Lambda^1M\otimes\g)$. A lift $M\ovs{u}\lra G$ is
admissible if $u^{-1}du\in\mathcal{E}(\Lambda^1M\otimes\g)$, a map
$M\ovs{\psi}\lra X$ is admissible if $\psi =u\pfi$ for a smooth $\pfi$
and an admissible $u$. We write $\mathcal{E}(M,G)$, $\mathcal{E}(M,X)$
for admissible lifts and maps respectively, and often shortly
$\mathcal{E}\pfi$ instead of $\mathcal{E}(M,G)\pfi$ for the
admissible $2$--homotopy sector of $\pfi$. 
\end{definition}
Despite the appearences our definition of spaces depends on a choice of $\pfi$ since $\bot$ stands for $\pr_{\h_\pfi^\bot}$. To avoid cumbersome symbols we often do not reflect this dependence in the notation assuming that a reference map is fixed once and for all. Note that conditions {\rm 1)}, {\rm 2)} of \eqref{admis} simply mean that $a$ has finite energy \eqref{e0.13}. In contrast, {\rm 3)} is stronger since in general one can only expect $a^{\|}\in
L^2(\Lambda^1M\otimes\g)$.  Unlike $W^{1,2}(M,G)$ the space $\mathcal{E}(M,G)$ is not a group. In fact, even if $u\in\mathcal{E}(M,G)$ and $v\in W^{2,2}(M,G)$ the product $uv$ may not be admissible. This is because
\be
(uv)^{-1}d(uv)^\bot=(\Ad_*(v^{-1})u^{-1}du)^\bot+(v^{-1}dv)^\bot
\ee
and $\Ad_*(v^{-1})$ does not commute with $\bot$, so the term
$((\Ad_*(v^{-1})u^{-1}du)^\bot)^{\wedge 2}$ may not be in $L^2$. 
However, if $w\in W^{2,2}(H_\pfi)$, i.e. {\it if in addition to $W^{2,2}$ regularity 
$w$ stabilizes $\pfi$ then $uw$ is again admissible}. Indeed, $E(uw\pfi)=E(u\pfi)<\infty$ guarantees
conditions {\rm 1)}, {\rm 2)} in \eqref{admis}. Also, $\Ad_*(w^{-1})$ commutes with $\pr_{\h_\pfi}$, $\pr_{\h_\pfi^\bot}$ when $w\pfi=\pfi$. Therefore,
$$
(\Ad_*(w^{-1})u^{-1}du)^{\|}=\Ad_*(w^{-1})(u^{-1}du)^{\|}
$$ 
and $(w^{-1}dw)^{\|}\in W^{1,2}(\Lambda^1M\otimes\g)$ so {\rm 3)} holds. In other words,
gauge-fixing by a $W^{2,2}$ transformation leaves us within the
class of admissible potentials. This will be crucial in the proof of \refT{3.1}.
\begin{definition}[weak convergence]\label{D:wconv} On $\mathcal{E}(M,G)$ define the natural weak convergence $u_n\ovs{\mathcal{E}}\rightharpoonup u$ by
\begin{equation}\label{e0.18}
\begin{aligned}
& {\rm 1)}\ u_n\overset{W^{1,2}}\lrhu u;\\
& {\rm 2)}\ a_n^{\perp}\wedge a_n^{\perp}\overset{L^2}\lrhu a^{\perp}\wedge a^{\perp};\\
& {\rm 3)}\ a_n^\vert\overset{W^{1,2}}\lrhu a^\vert,
\end{aligned}
\end{equation}
where of course $a_n=u_n^{-1}du_n$ and $a=u^{-1}du$.
\end{definition}
The space of admissible maps is obviously closed under both the weak and the strong convergence (obtained by replacing $\lrhu$ by $\to$ in \eqref{e0.18}). Our first observation is that the weak convergence behaves reasonably well with respect to multiplication. For two sequences of maps $u_n\ovs{\mathcal{E}}\rightharpoonup u$,
$v_n\ovs{\mathcal{E}}\rightharpoonup v$ does not necessarily imply
$u_nv_n\ovs{\mathcal{E}}\rightharpoonup uv$. As a matter of fact, 
$u_nv_n$ may not even belong to $\mathcal{E}(M,G)$. However,
\begin{lemma}\label{L:cow}
Let $u_n\ovs{\mathcal{E}}\rightharpoonup u$ and either
$w_n\ovs{C^\infty}\lra w$ or $w_n\in W^{2,2}(H_\pfi)$ and $w_n
\ovs{W^{2,2}}\lra w$. Then $u_nw_n\ovs{\mathcal{E}}\rightharpoonup
uw$.
\end{lemma}
\begin{proof}
$C^\infty$ case follows trivially from the definition. For the
$W^{2,2}$ case note that {\rm 2)} in \eqref{e0.18} can be replaced by
\bee\label{e0.112}
D_\pfi a_n\wedge D_\pfi
a_n\ovs{L^2}\rightharpoonup D_\pfi a\wedge D_\pfi a
\eee
with $D_\pfi a:=a^\bot+\pfi^*\omega^\bot$, see \eqref{e0.13}. The gain is that
for $a_n^{w_n}=(u_nw_n)^{-1}$ $d(u_nw_n)$ and $w_n\in W^{2,2}(H_\pfi)$
\bee\label{e0.113}
D_\pfi(a_n^{w_n})=\Ad_*(w_n^{-1})(D_\pfi a_n)\quad \text{a.e.}
\eee
Since $W^{2,2}(M,G)\subset C^0(M,G)$ by the Sobolev embedding theorems we have
$$
w_n\ovs{C^0}\lra w,\quad \Ad_*(w_n^{-1})\ovs{C^0}\lra \Ad_*(w^{-1})
$$
and therefore
\begin{multline*}
D_\pfi(a_n^{w_n})\wedge D_\pfi(a_n^{w_n})=\Ad_*(w_n^{-1})(D_\pfi a_n\wedge D_\pfi a_n)\\
\ovs{L^2}\rightharpoonup \Ad_*(w^{-1})(D_\pfi a\wedge D_\pfi a)=D_\pfi(a^w)\wedge D_\pfi(a^w).
\end{multline*}
The conditions {\rm 1)}, {\rm 3)} in \eqref{e0.18} can be checked similarly using \eqref{e0.113}
and the fact that $\Ad_*(w^{-1})$ commutes with $\pr_{\h_\pfi}$, $\pr_{\h_\pfi^\bot}$.
\end{proof}
We now wish to extend the description of homotopy classes from \refT{02} to admissible maps. Recall that the extra condition on $u$ is that $\int_M u^*\Theta\in\O_\pfi$, where $\Theta:=(\Theta_{\g_1},\dots,\Theta_{\g_N})$ and $\Theta_{\g_k}$  given by \eqref{e0.11} correspond to the decomposition $\g=\g_1\oplus\dots\oplus\g_N$ into simple components.
For brevity set $\alpha_{\g_k}:=\pr_{\g_k}(\alpha)$ for any $\g$--valued form $\alpha$.
Then \eqref{e0.11} implies for smooth maps
\bee\label{Thk}
u^*\Theta_{\g_k}:=c_{G_k}\tr((u^{-1}du)_{\g_k}\wedge(u^{-1}du)_{\g_k}\wedge(u^{-1}du)_{\g_k})=
c_{G_k}\tr(a_{\g_k}\wedge a_{\g_k}\wedge a_{\g_k}),
\eee
where as usual $a=u^{-1}du$. Note that the expression on the right is defined almost everywhere as a form even if $u$ is just a $W^{1,2}$ map.

It is easy to see from the product rule and the definition of Sobolev norms that
\be
||\alpha_{\g_k}||_{W^{l,p}}\leq ||\alpha||_{W^{l,p}}
\ee
for any form $\alpha$. Moreover, for any pair of forms $\alpha,\beta$
\be
(\alpha\wedge\beta)_{\g_k}=\alpha_{\g_k}\wedge\beta_{\g_k}
\ee
since elements from different $\g_k$ always commute. Therefore, if $a$ is admissible we have for each $k$:
\bee\label{kadmis}
\bal
{\rm 1)}\ &(a^\bot)_{\g_k} \in L^2(\Lambda^1M\otimes\g)\\
{\rm 2)}\ &(a^\bot)_{\g_k}\wedge(a^\bot)_{\g_k}=(a^\bot\wedge a^\bot)_{\g_k}\in L^2(\Lambda^2M\otimes\g)\\
{\rm 3)}\ &(a^\vert)_{\g_k}\in W^{1,2}(\Lambda^1M\otimes\g).
\eal
\eee
By the way, each $a_{\g_k}$ separately may not be admissible since in general
$(a_{\g_k})^\vert\neq(a^\vert)_{\g_k}$, $(a_{\g_k})^\bot\neq(a^\bot)_{\g_k}$. 

Even though $u^*\Theta$ is defined almost everywhere as a form in order 
to integrate it over $M$ we need it to be in $L^1$. Since we only know that $a_{\g_k}\in L^2$ the triple product $a_{\g_k}\wedge a_{\g_k}\wedge a_{\g_k}$ may not be integrable and one can not use expression \eqref{Thk} for integration directly. To take advantage of the conditions \eqref{kadmis} we decompose $a_{\g_k}=(a^\vert+a^\bot)_{\g_k}$,
plug it into $a_{\g_k}\wedge a_{\g_k}\wedge a_{\g_k}$ and use the distributive law. The resulting sum will have terms like $(a^\bot)_{\g_k}\wedge (a^\vert)_{\g_k}\wedge (a^\bot)_{\g_k}$, that are still not in $L^1$. Fortunately, we only have to integrate {\it traces} of such terms and the situation can be helped.
\begin{lemma}\label{L:HopInv}
Let $u\in\E(M,G)$ and $a=u^{-1}du$. Set $a^\vert_{\g_k}:=(a^\vert)_{\g_k}$, $a^\bot_{\g_k}:=(a^\bot)_{\g_k}$ and
\begin{multline}\label{traaa}
u^*\Theta_{\g_k} :=\tr(a^{\vert}_{\g_k}\wedge a^{\vert}_{\g_k}\wedge a^{\vert}_{\g_k})
+3\tr(a^{\vert}_{\g_k}\wedge a^{\vert}_{\g_k}\wedge a^{\bot}_{\g_k})\\
+3\tr(a^{\vert}_{\g_k}\wedge a^{\bot}_{\g_k}\wedge a^{\bot}_{\g_k})
+\tr(a^{\bot}_{\g_k}\wedge a^{\bot}_{\g_k}\wedge a^{\bot}_{\g_k}). 
\end{multline}
Then $u^*\Theta_{\g_k}\in L^1(\Lambda^3M)$ and is equal to the usual pullback if $u$ is smooth. 
\end{lemma}
\begin{proof}
Since $\tr(\xi_1\cdots\xi_n)$ is invariant under cyclic permutations of $\xi_k$-s we get for
any cyclic permutation $\sigma$ and $1$-forms $\alpha_k$:
\be
\tr(\alpha_{\sigma
(1)}\wedge\dots\wedge\alpha_{\sigma (n)})=(-1)^\sigma \mathop{\rm
tr}\nolimits(\alpha_1\wedge\dots\wedge\alpha_n)=(-1)^{n-1}\mathop{\rm
tr}\nolimits(\alpha_1\wedge\dots\wedge\alpha_n).
\ee 
As a corollary for any smooth forms $\alpha$, $\beta$ the wedge cube $\tr((\alpha +\beta)^{\wedge 3})$ 
reduces to the binomial form 
\be 
\tr((\alpha +\beta)^{\wedge 3})
=\tr(\alpha^{\wedge 3})
+3\tr(\alpha^{\wedge 2}\wedge\beta)
+3(\alpha\wedge\beta^{\wedge 2})
+\tr(\beta^{\wedge 3}).
\ee
Applying it to $\alpha=a^\vert_{\g_k}$, $\beta=a^\bot_{\g_k}$ we see that $\tr(a_{\g_k}\wedge a_{\g_k}\wedge a_{\g_k})$ is equal to the righthand side of \eqref{traaa}. When $u$ and hence $a$ are admissible one derives from \eqref{kadmis} and the Sobolev multiplication theorems 
\bee\label{trspace}
\bal
{\rm 1)}\ &a^{\vert}_{\g_k}\wedge a^{\vert}_{\g_k}\wedge a^{\vert}_{\g_k}\in L^2\\
{\rm 2)}\ &a^{\vert}_{\g_k}\wedge a^{\vert}_{\g_k}\wedge a^{\bot}_{\g_k}\in L^{6/5}\\
{\rm 3)}\ &a^{\vert}_{\g_k}\wedge a^{\bot}_{\g_k}\wedge a^{\bot}_{\g_k}\in L^{3/2}\\
{\rm 4)}\ &a^{\bot}_{\g_k}\wedge a^{\bot}_{\g_k}\wedge a^{\bot}_{\g_k}\in L^1.
\eal
\eee
Now by Sobolev embeddings $u^*\Theta_{\g_k}\in L^1(\Lambda^3M)$.
\end{proof}
If we knew only that $a^{\vert}_{\g_k}\in L^2$ then the first two terms in \eqref{trspace} may not be in $L^1$. This justifies the introduction of admissible maps. In some cases however, one can do without them. For example, if $G$ is a simple group and the subgroup $H$ is Abelian one has $[\h,\h]=0$ and hence $a^{\vert}\wedge a^{\vert}=0$, so the singular terms vanish. This is the case if $X=SU_2/U_1$ or more generally, a flag manifold $X=SU_{n+1}/\T^n$, where $\T^n$ is a maximal torus.
\begin{definition}[Homotopy sector]\label{D:hsector}
An element $\psi\in\mathcal{E}(M,X)$ is in the homotopy sector $\mathcal{E}_\pfi$ of
$\pfi$ if
\bee\label{e0.125}
\bal
&{\rm 1)}\ \psi=u\pfi\quad \text{with}\quad u\in\mathcal{E}(M,G)\\
&{\rm 2)}\ \int\limits_Mu^*\Theta=0\mod\O_\pfi,
\eal
\eee
where $u^*\Theta:=(u^*\Theta_{\g_1},\dots,u^*\Theta_{\g_N})$ is defined by \eqref{traaa}.
\end{definition}
\noindent If $\psi\in C^1(M,X)$ then by \refT{02} it is in the homotopy sector of
$\pfi$ if and only if $\psi$ is homotopic to $\pfi$ in the usual sense.

Even though the integral $\int_M u^*\Theta$ is now defined for all admissible maps it may not behave well under weak convergence. Given $u_n\ovs{\E}\rightharpoonup u$ we need $u^*(\Theta_n)_{\g_k}\ovs{\mathcal{D}'}\rightharpoonup u^*\Theta_{\g_k}$ in the space of {\it Schwarz distributions} $\mathcal{D}'$ to have the integrals converge. As usual, $\mathcal{D}$ is the space of test forms, $C^\infty$ with compact support, and $\mathcal{D}'$ is the dual space relative to the inner product in $L^2$ \cite{GMS1}. The first three terms in \eqref{traaa} trivially converge even in $L^1$. 
Therefore, we just need 
$$
\tr\left((a_n^\bot)_{\g_k}\wedge(a_n^\bot)_{\g_k}\wedge(a_n^\bot)_{\g_k}\right)\ovs{\mathcal{D}'}
\rightharpoonup\tr\left((a^\bot)_{\g_k}\wedge(a^\bot)_{\g_k}\wedge(a^\bot)_{\g_k}\right).
$$
Distributional convergence of wedge products is a well-studied subject and we now recall a relevant result from \cite{RRT} (see also \cite{IV} for a different approach). 
\begin{theorem*}[Wedge Product theorem, \cite{RRT}] 
Assume that $\upsilon_n\ovs{L^2}\rightharpoonup\upsilon$,
$\omega_n\ovs{L^2}\rightharpoonup\omega$ are sequences of $L^2$ differential
forms on a compact manifold $M$ and $d\upsilon_n$, $d\omega_n$ are precompact in $W^{-1,2}$.
Then $\upsilon_n\wedge\omega_n\ovs{\mathcal{D}'}\rightharpoonup\upsilon\wedge\omega$.
\end{theorem*}
\noindent It will be convenient for us to use the Wedge Product theorem in a slightly weakened form. By a Sobolev embedding theorem $L^s\hra W^{-1,p}$ compactly if $\frac{1}{s}<\frac{1}{n}+\frac{1}{p}$ ($n:=\dim\,M$). For a 
$3$-dimensional $M$ and $p=2$ this gives $s>\frac{6}{5}$. Thus, we can replace precompactness in $W^{-1,2}$ by boundedness in $L^{6/5+\ve}$ with $\ve>0$.

Even with the Wedge Product theorem we are unable to prove convergence for general homogeneous spaces. The next lemma requires extra cancelations that happen in symmetric spaces (see the discussion after \refC{refcurv} and \cite{Ar,Hl}).
\begin{lemma}\label{L:sconv}
If $X$ is a Riemannian symmetric space then $u_n\ovs{\mathcal{E}}\rightharpoonup u$ implies
$u_n^*\Theta\ovs{\mathcal{D}'}\rightharpoonup u^*\Theta$ and therefore
$$
\int\limits_Mu_n^*\Theta\to \int\limits_Mu^*\Theta.
$$
\end{lemma}
\begin{proof}
By the Wedge Product theorem it suffices to show that $d(a_n^\bot\wedge a_n^\bot)$ is bounded in $L^{6/5+\ve}$.
The first term on the right of \eqref{daa} is manifestly in $L^{3/2}$. Now recall the cancellation formula $[[\alpha ,\beta], \beta]=[\alpha ,\beta\wedge\beta]$ that holds for all $\g$-valued forms $\alpha$ and odd degree forms $\beta$. Applying it to the third term,
\be
[[a^{\|},a^\bot],a^\bot]=[a^{\|},a^\bot\wedge a^\bot]\in L^{3/2}.
\ee
For the second and the fourth terms in general we only have
$$
[d\Phi\wedge a^\bot,a^\bot]\in L^1\quad \text{and}\quad [(I-\Phi)(a^\bot\wedge a^\bot),a^\bot]\in L^1,
$$ 
while $1<6/5$. But if $X$ is symmetric then the fourth term vanishes altogether and the sum of the second and the third is even in $L^2$, see \refL{dcurv}${\rm (iii')}$.
\end{proof}
There are more properties that $\int\limits_Mu^*\Theta$ should have to qualify as a topological degree \cite{BT}. For one, it should only take {\it integral values} as it does on smooth maps. Moreover, for smooth maps this integral is a group homomorphism \cite{Dy}, i.e.
\bee\label{additv}
\int\limits_M(uv)^*\Theta=\int\limits_Mu^*\Theta+\int\limits_Mv^*\Theta.
\eee
One can not expect \eqref{additv} to hold when both $u,v$ are just admissible since the lefthand side may
not be defined. But even assuming that $v$ is smooth it is unclear if \eqref{additv} holds for all admissible $u$.
The underlying difficulty is that we do not know if one can approximate an admissible $u$ by smooth maps of the same degree. This gives a rationale for introducing the strongly admissible maps next.
\begin{definition}[strongly admissible maps]\label{D:sadmis}
Denote $\E'(M,G)$ the sequentially weak closure of $C^\infty(M,G)$ in $\E(M,G)$. Spaces $\mathcal{E'}(\Lambda^1M\otimes\g)$ and $\mathcal{E'}(M,X)$ are defined analogously to \refD{admis}, and $\mathcal{E'}\pfi$, $\mathcal{E}'_\pfi$ denote the $2$-homotopy sector and the  homotopy sector of $\pfi$ respectively.
\end{definition}
\noindent Similarly constructed spaces have been used in \cite{Es1,GMS1} for similar problems. It may well be that 
$\mathcal{E}(M,X)=\mathcal{E}'(M,X)$, but the question is still open even for $X=SU_2$ (see \cite{Es2}). From this definition we can only claim that $W^{2,2}(M,G)\subset\mathcal{E}'(M,G)$ because $W^{2,2}(M,G)\subset C^0(M,G)$ by the Sobolev embedding theorems \cite{Pl}. In fact, it is contained even in the strong closure of $C^\infty$ in $\mathcal{E}$.

The next Lemma shows that strongly admissible maps on symmetric spaces share many topological properties with smooth maps.
\begin{lemma}\label{L:sadmis}
Let $X=G/H$ be a Riemannian symmetric space and $M\ovs{\pfi}\lra G$ be a smooth
reference map. Then

\noindent {\rm (i)\bf(integrality)} For a strongly admissible map $u\in\mathcal{E}'(M,G)$ the degree is integral:
$$
\int\limits_Mu^*\Theta\in\Z^N.
$$
\noindent {\rm (ii)\bf(stabilizer)} If $w\in W^{2,2}(M,G)$ stabilizes $\pfi$, i.e. $w\in W^{2,2}(H_\pfi)$ then
$$
\int\limits_Mw^*\Theta=0\mod\O_\pfi
$$
\noindent {\rm (iii)\bf(additivity)} If $u\in\mathcal{E}'(M,G)$ and either $w\in
C^\infty(M,G)$ or $w\in W^{2,2}(H_\pfi)$ then $uw\in\mathcal{E}'(M,G)$ and
\bee\label{e0.119}
\int\limits_M(uw)^*\Theta=\int\limits_M u^*\Theta+\int\limits_M w^*\Theta
\eee
\noindent {\rm (iv)\bf(smooth representative)}
Every homotopy sector of strongly admissible maps contains a smooth representative.

\noindent {\rm (v)\bf(change of reference)} If two smooth maps are homotopic they define the same homotopy sector of strongly admissible maps.
\end{lemma}
\begin{proof}
{\rm (i)-(iii)} follow by smooth approximation in view of \refL{sconv}.
\bigskip

\noindent {\rm (iv)} By definition of $\mathcal{E}'(M,X)$ for any map $\psi\in
C^\infty(M,X)$ there is $\widetilde{\pfi}\in C^\infty(M,X)$ and $\widetilde{u}\in\mathcal{E}'(M,G)$ with
$\psi=\widetilde{u}\widetilde{\pfi}$. Then the vector
$$
\nu:=\int\limits_M\widetilde{u}^*\Theta
$$ 
is in $\Z^N$ by (i). By the Eilenberg classification theorem \cite{K,St} there is a $v\in C^\infty(M,G)$
such that 
$$
\int\limits_Mv^*\Theta=\nu.
$$ 
Set $u:=\widetilde{u}v^{-1}$, $\pfi:=v\widetilde{\pfi}$ then still $\psi=u\pfi$. 
By (iii) $u\in\mathcal{E}'(M,G)$ and
\be
\int\limits_Mu^*\Theta=\int\limits_M\widetilde{u}^*\Theta-\int\limits_Mv^*\Theta=0
\ee
so $\psi\in\mathcal{E}_\pfi'$, where $\pfi$ is smooth by construction.
\bigskip

\noindent {\rm (v)} Let $\widetilde{\pfi},\pfi$ be smooth and homotopic.
It follows from \refT{02} that $\O_{\widetilde{\pfi}}=\O_\pfi$ and there is a smooth $v$ such that $\widetilde{\pfi}:=v\pfi$. Moreover, $v$ can be chosen nullhomotopic so that $\int\limits_Mv^*\Theta=0$.
Let $\psi=u\pfi\in\mathcal{E}_\pfi'$ be arbitrary. By definition of $\mathcal{E}_\pfi'$ we have $\int\limits_Mu^*\Theta=0\mod\O_\pfi$. Set $\widetilde{u}:=uv^{-1}$ then 
$\psi=\widetilde{u}\widetilde{\pfi}$ and by (iii):
\be
\int\limits_M\widetilde{u}^*\Theta
=\int\limits_Mu^*\Theta+\int\limits_M(v^{-1})^*\Theta
=\int\limits_Mu^*\Theta-\int\limits_Mv^*\Theta=0\mod\O_\pfi=\O_{\widetilde{\pfi}}.
\ee
Thus, $\psi\in\mathcal{E}_{\widetilde{\pfi}}'$ and $\mathcal{E}_\pfi'\subset\mathcal{E}_{\widetilde{\pfi}}'$. The other inclusion follows by switching $\pfi$ and $\widetilde{\pfi}$.
\end{proof}
Thus, strongly admissible maps on symmetric spaces are topologically reasonable 
and at the same time, closed under weak limits. This makes them particularly suitable for solving variational problems. It may even be argued (see \cite{GMS1}) that this class is more natural than $\mathcal{E}(M,X)$ since we really want to minimize energy over smooth maps. The restriction to symmetric spaces is unfortunate, but it appears to be the natural generality of our approach.

\section{Gauge fixing and minimization}\label{S5}

In this section we prove our main results on existence of Hopfions. We give a complete proof for the case of Riemannian symmetric spaces and prove a weaker result in the general case. Unlike in the case of maps problems with smooth approximation do not arise for differential forms since their spaces are linear. Hence, the formulas derived in \refL{dcurv} for $C^\infty$ potentials still hold for admissible ones in the distributional sense.

In particular, by smooth approximation of $u$ in $W^{1,2}(M,G)$ pure-gauge admissible potentials $a=u^{-1}du$ satisfy 
$$
da+a\wedge a=0\quad \text{(equality in $W^{-1,2}(\Lambda^2M\otimes\g)),$}
$$
i.e. they are distributionally flat. Note that for $a$ in $L^2$ the relation
$da=-a\wedge a$ implies that $da$ which is a priori only in
$W^{-1,2}$ is actually in $L^1$. If moreover $a$ has finite energy \eqref{e0.13}, then \eqref{dcurv} yields 
$$
F(a^{\|})\in L^2\quad \text{and}\quad (da^\bot)^{\|}\in L^2.
$$
The other component $(da^\bot)^\bot$ is spoiled by the term $[a^{\|},a^\bot]$ which
will only be in $L^{3/2}$ even assuming that $a$ is admissible,
i.e. $a^{\|}\in W^{1,2}$.

Let us say a few words about the role the gauge theory plays in
the proofs. When we attempt to minimize (\ref{e0.6}) the following problem presents itself. Given $\psi$ and $\pfi$ 
the choice of $u$ in $\psi=u\pfi$ is not unique. Without changing $\psi$ it can be
replaced by $uw$, where $w$ stabilizes $\pfi$, i.e. $w\pfi=\pfi$. Since the functional (\ref{e0.6}) only depends on
$\psi$ it remains invariant under this change and therefore admits a
non-compact group of symmetries as a functional \eqref{e0.13} of $u$ or $a$. As
a result, sets of bounded energy are not weakly compact in
any reasonable sense. This sort of malaise is well known in 
gauge theory, where the group of symmetries is the gauge group of a
principal bundle acting on connections. A cure is to fix the gauge.

As shown in \refS{S3} the isotropic part
$a^\vert:=\pr_{\h_\pfi}(a)$  gives the gauge potential of a
connection on the subbundle $\pfi^*P\subset M\times G$ under the
identification of \refL{2.3}. Moreover, if $u$ is replaced
by $uw$ and hence $a$ is replaced by $a^w:=(uw)^{-1}d(uw)$ then
$(a^w)^\vert=(a^\vert)^w$, where on the right we have the expression
from \eqref{e0.60}(i). In other words, as far as the isotropic parts
are concerned {\it the action of $\pfi$-stabilizing maps is
conjugate to the action of the gauge group $\Gamma(\Ad(\pfi^*P))$
on connections}. \refT{2.1}(iii) along with the flatness of $a$
implies that
\begin{equation}\label{e0.19}
F(a^\vert)=d(\pr_{\h_\pfi})\wedge a^\perp-(a^\perp\wedge a^\perp)^\vert
-(\pfi^*\omega^\perp\wedge\pfi^*\omega^\perp)^\vert
\end{equation}
and $a^\perp$, $a^\perp\wedge a^\perp$ are bounded in $L^2$ by the
functional (\ref{e0.13}). This is a key relation connecting topology of maps 
to the Faddeev-Skyrme functional. The Uhlenbeck compactness theorem below implies then that
$a^\vert$ can be controlled by fixing the gauge in $\Ad_*(\pfi^*P)$. In terms of maps this means that we replace $u$
by a suitable $uw$ when representing $\psi$ in the minimization process.
\begin{theorem*}[Uhlenbeck compactness theorem, \cite{Ul1,We}] 
Let $P\to M$ be a smooth principal bundle and $2p>\mbox{dim}M$. Consider a sequence of gauge
potentials on $M$
$$
\alpha_n\in W^{1,p}(\Lambda^1M\otimes\Ad_*P)\quad \text{with}\quad ||F(\alpha_n)||_{L_p}\leq C<\infty.
$$  
Then there exists a subsequence $\alpha_{n_k}$ along with gauge transformations $\lambda_{n_k}\in W^{2,p}(\Ad P)$ such that
\bee\label{e0.106}
\alpha_{n_k}^{\lambda_{n_k}}\ovs{W^{1,p}}\rightharpoonup\alpha\qquad
\mbox{and}\quad ||F(\alpha)||_{L_p}\leq C.
\eee
\end{theorem*}
\begin{remark*} Note that in the Uhlenbeck compactness theorem $\alpha_n$ 
are assumed from the start to be in $W^{1,p}$ rather than just in $L^p$. 
It is an open question if one could assume in the Uhlenbeck theorem 
$\alpha_n\in L^p(\Lambda^1M\otimes\Ad_*P)$ while allowing $\lambda_{n_k}\in W^{1,p}(\Ad P)$. 
One can show that this is the case at least when the gauge group is Abelian.
\end{remark*}
We will use this compactness theorem to fix the gauge for the isotropic parts $a_n^{\|}$
of potentials in a minimizing sequence. This means that we need $a_n^{\|}\in
W^{1,2}(\Lambda^1M\otimes\g)$ from the start to apply the theorem, and this is another reason for restricting to the admissible maps. 

We can rewrite \eqref{e0.6} as
\bee\label{FadSkyr}
E(\psi)=\int\limits_M\frac{1}{2}|\psi^*\omega^\bot|^2+\frac{1}{4}|\psi^*\omega^\bot\wedge\psi^*\omega^\bot|^2\,dm.
\eee
Evaluated on $\psi=u\pfi$ it becomes
\begin{equation}\label{FadSkyrp}
E_\pfi(a)=\int_M\frac12|D_\pfi a|^2\, +\,\frac14|D_\pfi a\wedge
D_\pfi a|^2\;dm.
\end{equation}
with notations $a=u^{-1}du$ and $D_\pfi a:=a^\perp+\pfi^*\omega^\perp$.
\begin{theorem}\label{T:3.1}
Every $2$--homotopy sector of admissible maps has a minimizer of the Faddeev-Skyrme
energy.
\end{theorem}
\begin{proof}
We denote by $\ovs{L}\rightharpoonup$
($\ovs{L}\lra$) the weak (the strong) convergence in a Banach space $L$. 
All constants in the estimates are denoted by $C$ even though they may be different. 
Passing to subsequences is also ignored in the notation. This does not lead to any confusion. 

Recall that we assume $G\hra\mbox{End}(\mathbb{E})$ for a Euclidean space $\mathbb{E}$ and
$u\in W^{1,2}(M,G)$ means $u\in W^{1,2}(M,\mbox{End}(\mathbb{E}))$ with $u(m)\in G$ a.e. 
Let $\psi_n=u_n\pfi$ be a minimizing sequence of admissible maps in a sector $\mathcal{E}\pfi$ and 
$a_n:=u_n^{-1}du_n$. The proof is divided into several steps.

\noindent {\bf Gauge-fixing}\nopagebreak 

\noindent By definition
$$
E(u_n\pfi)=E_\pfi(a_n)\leq C<\infty.
$$
It follows by inspection from \eqref{FadSkyrp} that 
$$
||a_n^\bot||_{L^2}\leq C<\infty\quad \text{and}\quad 
||a_n^\bot\wedge a_n^\bot||_{L^2}\leq C<\infty.
$$ 
Then by \refL{dcurv}(i) also
$$
||F(a_n^{\|})||_{L^2}\leq C<\infty.
$$
Since $u_n$ are admissible $a_n^{\|}\in W^{1,2}$ 
and we may apply the Uhlenbeck compactness theorem to $a_n^{\|}$. After passing to a
subsequence we get a sequence of gauge transformations $w_n\in W^{2,2}(H_\pfi)$ such that
$$
(a_n^{\|})^{w_n}=(a_n^{w_n})^{\|}\ovs{W^{1,2}}\rightharpoonup
a^{\|}.
$$
But 
$$
a_n^{w_n}=\Ad_*(w_n^{-1})a_n+w_n^{-1}dw_n=(u_nw_n)^{-1}d(u_nw_n)
$$
and $u_nw_n$ are still admissible. Therefore we can drop $w_n$ from the notation 
and assume that $u_n$ are preselected to have the isotropic components $a_n^{\|}$ 
weakly convergent in $W^{1,2}$.

\noindent {\bf Compactness} 

\noindent Let $u_n$ be the gauge-fixed minimizing sequence from the previous step.
Since $G$ is compact it is bounded in $\End(\mathbb{E})$ and
$$
\|u_n\|_{L^\infty}\leq C<\infty.
$$
By gauge-fixing and \eqref{FadSkyrp} both
$a_n^{\|}$, $a_n^\bot$ are bounded in $L^2$. Therefore, so are 
$$
a_n=a_n^{\|}+a_n^\bot=u_n^{-1}du_n\quad \text{and}\quad du_n=u_na_n.
$$ 
We conclude that
$$
\|u_n\|_{W^{1,2}}\leq C<\infty
$$ 
and after passing to a subsequence $u_n\ovs{W^{1,2}}\rightharpoonup u$. 

Since $W^{1,2}\hra L^2$ is a compact embedding we have $u_n\ovs{L^2}\lra u$ and since
$u_n$ are bounded in $L^\infty$ also $u_n^{-1}\ovs{L^2}\lra u^{-1}$. But the strong
convergence in $L^2$ implies convergence almost everywhere on a subsequence and we have $u(m)\in G$ a.e. 
so that $u\in W^{1,2}(M,G)$. 

The differential $d:W^{1,2}\to L^2$ is a bounded linear operator 
and hence it is weakly continuous. Therefore
$$
du_n\ovs{L^2}\rightharpoonup du\quad \text{and}\quad
u_n^{-1}du_n=a_n\ovs{L^2}\rightharpoonup a:=u^{-1}du.
$$ 
Moreover, by the preselection of $u_n$ we have in addition 
$$
a_n^{\|}\ovs{W^{1,2}}\rightharpoonup a^{\|}\in W^{1,2}(\Lambda^1M\otimes\g).
$$

\noindent {\bf Closure} 

\noindent In view of \eqref{FadSkyrp} 
$$
\|a_n^\bot\wedge a_n^\bot\|_{L^2}\leq C<\infty
$$ 
and possibly after passing to another subsequence,
$$
a_n^\bot\wedge a_n^\bot\ovs{L^2}\rightharpoonup\Lambda.
$$ 
Since $a_n^\bot$ is bounded in $L^2$ and $a_n^{\|}$ is bounded in
$W^{1,2}$ we have by the Sobolev multiplication theorem \cite{Pl}:
$$
\|[a_n^{\|},a_n^\bot]\|_{L^{3/2}}\leq C<\infty
$$
and hence by \refL{dcurv}
$$
\|da_n^\bot\|_{L^{3/2}}\leq C<\infty.
$$ 
But $3/2>6/5$ and the Wedge Product theorem now implies 
$$
a_n^\bot\wedge a_n^\bot\ovs{\mathcal{D}'}\rightharpoonup a^\bot\wedge a^\bot.
$$ 
By uniqueness of the limit in $\mathcal{D}'$ one must have $\Lambda=a^\bot\wedge a^\bot$ and
$$
a_n^\bot\wedge a_n^\bot\ovs{L^2}\rightharpoonup a^\bot\wedge
a^\bot\in L^2(\Lambda^2M\otimes\g).
$$ 
Along with the previous step this yields $u\in\mathcal{E}(M,G)$ and hence
$\psi:=u\pfi\in\mathcal{E}(M,X)$. This is the map we were looking for.

\noindent {\bf Lower semicontinuity} 

\noindent $E$ in \eqref{FadSkyr} is not a weakly lower semicontinuous functional of $\psi$ 
and neither is $E_\pfi$ in \eqref{FadSkyrp} as a functional of $a$. 
However,
$$
\widehat{E}(r,\Lambda):=\frac{1}{2}\|r\|_{L^2}^2+\frac{1}{4}\|\Lambda\|_{L^2}^2
$$
{\it is} a weakly lower semicontinuous functional of a pair (see \cite{BlM}):
$$
(r,\Lambda)\in
L^2(\Lambda^1M\otimes \g)\times L^2(\Lambda^2M\otimes\g)
$$ 
But obviously, $E_\pfi(a)=\widehat{E}(D_\pfi a,D_\pfi a\wedge D_\pfi a)$.
By the above,
$$
D_\pfi a_n=\pfi^*\omega^\bot+a_n^\bot\ovs{L^2}\rightharpoonup D_\pfi a\quad \text{and}\quad
D_\pfi a_n\wedge D_\pfi a_n\ovs{L^2}\rightharpoonup D_\pfi a\wedge D_\pfi a.
$$ 
Therefore,
\begin{multline*}
E(\psi)=E_\pfi(a)=\widehat{E}(D_\pfi a,D_\pfi a\wedge D_\pfi a)\\
\leq \liminf_{n\to\infty}E(D_\pfi a_n,D_\pfi a_n\wedge D_\pfi a_n)
=\liminf_{n\to\infty}E_\pfi(a_n)=\liminf_{n\to\infty}E(\psi_n).
\end{multline*}
Since $\psi_n$ was a minimizing sequence in $\mathcal{E}\pfi$ and $\psi=u\pfi\in\mathcal{E}\pfi$ it is a minimizer
of \eqref{FadSkyr} in the $2$--homotopy sector of $\pfi$.
\end{proof}
\begin{remark*}
If $a_n$ are not  just flat but pure-gauge it follows from a result
in \cite{AK1} that on a subsequence $a_n\ovs{L^2}\rightharpoonup a$,
where $a$ is also pure-gauge. Using this result one could prove \refT{3.1}
without introducing $u_n$ explicitly, but such a proof requires a lengthy
discussion of holonomy for distributional connections.
\end{remark*}

For $X=S^2$ \refT{3.1} is proved in \cite{AK2} (Theorem 4). In fact the
result there is stronger: $\mathcal{E}\pfi$ is subdivided into subsectors by
additional Chern-Simons invariants and there is a separate minimizer
in each subsector. This already shows that a minimizer in
$\mathcal{E}\pfi$ is not unique. But even if $\pi_3(X)=0$ and the
$2$--homotopy sectors characterize homotopy classes completely there is little
hope that the minimizers of \eqref{FadSkyr} are unique since the functional is
nowhere near being convex. We now extend the $S^2$ result to all symmetric spaces.
\begin{theorem}\label{T:11}
Let $X$ be a symmetric space. Then every homotopy sector of strongly
admissible maps contains a minimizer of Faddeev-Skyrme energy.
\end{theorem}
\begin{proof}
We proceed as in the proof of \refT{3.1} by choosing a
minimizing sequence $\psi_n=u_n\pfi$, $u_n\in\mathcal{E}'(M,G)$ and
$\int\limits_Mu_n^*\Theta\in\O_\pfi$. Gauge-fixing replaces
$u_n$ by $u_nw_n$ with $w_n\in W^{2,2}(H_\pfi)$ and by \refL{sadmis}(ii),(iii)
\be
\int\limits_M(u_nw_n)^*\Theta=\int\limits_Mu_n^*\Theta+\int\limits_Mw_n^*\Theta=0\mod\O_\pfi,
\ee
i.e. we may assume having $u_nw_n$ from the start
and drop $w_n$ from the notation. Now setting $a_n=u_n^{-1}du_n$ we have
$a_n^{\|}\ovs{W^{1,2}}\rightharpoonup a^{\|}$ since $u_n$ is
gauge-fixed. As in the proof of primary minimization we establish on
a subsequence
\be
\bal
&u_n\ovs{W^{1,2}}\rightharpoonup u\\
&a_n^\bot\ovs{L^2}\rightharpoonup a^\bot\\
&a_n^\bot\wedge a_n^\bot\ovs{L^2}\rightharpoonup a^\bot\wedge a^\bot,
\eal
\ee
where $a:=u^{-1}du$. But this means that
$u_n\ovs{\mathcal{E}}\rightharpoonup u$ and by \refL{sconv}
\be
u_n^*\Theta\ovs{\mathcal{D}'}\rightharpoonup u^*\Theta,
\ee
i.e. $\int\limits_Mu^*\Theta\in  \O_\pfi$. Since $u$ is a limit in
$\mathcal{E}$ of maps from $\mathcal{E}'$ it is in $\mathcal{E}'$
itself and hence $\psi=u\pfi\in\mathcal{E}_\pfi '$. As in the proof of \refT{3.1}
$$
E(\psi)\leq\liminf_{n\to\infty}E(\psi_n)
$$ 
and since $\psi_n$ was a minimizing sequence $\psi$ is a minimizer in $\mathcal{E}_\pfi'$.
\end{proof}

\section*{Open problems}

Although generalized Hopf invariants are probably ill-behaved on maps to general codomains, we believe that our methods generalize at least to flag manifolds $G/\T$ ($\T$ is a maximal torus of a Lie group $G$). Flag manifolds appear in the Faddeev-Niemi conjecture \cite{FN2}, which states that the $SU_{n}/\T$ Faddeev-Skyrme model describes a low-energy limit of the $SU_{n}$ Yang-Mills theory. This is supported by the fact that $SU_{n}/\T$ Hopfions can be lifted to stationary points of $SU_{n}$ Skyrme model, whose connection to the Yang-Mills theory is well-established \cite{Ch}. Stability of lifted Hopfions under the Skyrme functional is an interesting open question.

The Faddeev-Niemi conjecture motivates studying the topology of the configuration spaces of the $SU_{n}/\T$ Faddeev-Skyrme models and comparing it to the topology of the Yang-Mills configuration space. For the case of the $2$--sphere the fundamental group and the real cohomology ring of the configuration space were computed in \cite{AS}. It is instructive to generalize the computation to flag manifolds. It is also interesting to explore recent {\it gauged Skyrme} models, where extra gauge fields are present along with additional terms in energy functionals \cite{HZ,NSK}. Some of them are predicted to have self-duality properties similarly to pure Yang-Mills fields.

A challenging problem is to replace closed $3$-manifolds as domains of maps. Whereas the results of this paper generalize to bounded domains in $\R^3$ rather straightforwardly, it is not the case with non-compact manifolds, unbounded domains in $\R^3$ or even $\R^3$ itself. The case of $\R^3$ is the most natural from the physical point of view. As suggested by \cite{KV,LY2} an important step is to obtain an asymptotic growth estimate for energy of minimizers as a function of their topological numbers (degree, Hopf invariant, etc.). We know that the growth is linear for Lie groups and fractional with power $3/4$ for $SU_2/U_1$, a similar estimate was proved in \cite{Sh2} for the Faddeev-Niemi functional.

The best result so far for $\R^3$ only gives existence of infinite number of Hopfions with unknown Hopf invariants \cite{LY2}. The minimization problem on $\R^3$ has a specific difficulty of maps jumping from one homotopy class to another in the limit due to effects at infinity. On the other hand, the Uhlenbeck compactness theorem has been recently generalized to some non-compact manifolds in \cite{We}. Hopefully, the gauge methods of this work combined with these new results will lead to a complete solution for $\R^3$.

It is interesting that for bounded domains there is a linear lower bound on energy even if the Dirichlet term is dropped \cite{CDG}. One would like to find analogous growth estimates for other homogeneous spaces $G/H$ and investigate the dependence of the power of the growth on a way $H$ sits inside of $G$ for both bounded and unbounded domains. 

Finally, regularity and fine geometry of Hopfions remain widely open even in the Faddeeev model. The conjecture is that they are smooth, but no path towards a proof has emerged so far. It is equally unclear how to identify which links appear as soliton cores at different values of the Hopf invariant.

{
\renewcommand{\baselinestretch}{1}

\addcontentsline{toc}{chapter}{Bibliography}

\begin{thebibliography}{1000}


\footnotesize


\bibitem[Ar]{Ar} Arvanitoyeorges A.: An introduction to Lie groups and the geometry of homogeneous spaces. Student Mathematical Library,\ \textbf{22}, American Mathematical Society, Providence, RI, 2003.

\bibitem[AK1]{AK1} Auckly, D., Kapitanski, L.: Holonomy and Skyrme's model.
Comm. Math. Phys.,\ \textbf{240}(2003), no. 1-2, 97--122.

\bibitem[AK2]{AK2} Auckly, D., Kapitanski, L.: Analysis of $S^2$-valued maps and Faddeev's model. Comm. Math. Phys. ,\ \textbf{256}(2005), no. 3, 611--620.

\bibitem[AK3]{AK3} Auckly, D., Kapitanski, L.: The Pontrjagin-Hopf invariants for Sobolev maps. \texttt{arxiv:} math-ph/0711.0546

\bibitem[AS]{AS} Auckly, D., Speight M.: Fermionic quantization and configuration spaces for the Skyrme and Faddeev-Hopf models. Comm. Math. Phys.,\ \textbf{263}(2006),  no. 1, 173--216.

\bibitem[BMSS]{BMSS} Balachandran A., Marmo G., Skagerstam B., Stern A.:  Classical topology and quantum states. World Scientific Publishing Co., Inc., River Edge, NJ, 1991.

\bibitem[BlM]{BlM} Ball J., Murat F.: $W\sp{1,p}$-quasiconvexity and variational problems for multiple integrals. J. Funct. Anal.,\ \textbf{58}(1984), no. 3, 225--253.

\bibitem[BS1]{BS1} Battye R., Sutcliffe P.:  Solitons, links and knots.  R. Soc. Lond. Proc. Ser. A Math. Phys. Eng. Sci.,\ \textbf{455}(1999),  no. 1992, 4305--4331.

\bibitem[BT]{BT} Bott R., Tu L.:  Differential forms in algebraic topology. Graduate Texts in Mathematics,\ \textbf{82}, Springer-Verlag, New York-Berlin, 1982.

\bibitem[BtD]{BtD} Br\"oker T., tom Dieck T.: Representations of compact Lie groups. Graduate Texts in Mathematics,\ \textbf{98}, Springer-Verlag, New York, 1985.

\bibitem[CDG]{CDG} Cantarella J., DeTurck D., Gluck H.: Upper bounds for the writhing of knots and the helicity of vector fields, in \textit{Knots, braids, and mapping class groups}, 1--21, Amer. Math. Soc., Providence, RI, 2001. 

\bibitem[Ch]{Ch} Cho Y.: Reinterpretation of Faddeev-Niemi knot in Skyrme theory. Phys. Lett. B,\ \textbf{603}(2004), no. 1-2, 88--93. 

\bibitem[DFN]{DFN} Dubrovin B., Fomenko A., Novikov S.: Modern geometry---methods and applications, part II. The geometry and topology of manifolds. Graduate Texts in Mathematics,\ \textbf{104}, Springer-Verlag, New York, 1985.

\bibitem[Dy]{Dy} Dynkin E.:  Homologies of compact Lie groups.
Amer. Math. Soc. Transl. (2),\ \textbf{12}(1959), 251--300.

\bibitem[EL]{EL} Eells, J., Lemaire, L.: Selected topics in harmonic maps. AMS, Providence, 1983.

\bibitem[Es1]{Es1} Esteban M.: A direct variational approach to Skyrme's model for meson fields. Comm. Math. Phys.,\ \textbf{105}(1986), no. 4, 571--591.

\bibitem[Es2]{Es2} Esteban M.:  A new setting for Skyrme's problem, in \textit{Progress in Nonlinear Differential Equations and Their Applications}, vol. 4, 77--93, Basel-Boston, Birkh�ser, 1990.

\bibitem[Es3]{Es3} Esteban M.: Erratum: Existence of $3D$ Skyrmions. Complete version (2004). math-ph/0401042.

\bibitem[Fd1]{Fd1} Faddeev L.: Quantization of solitons.
[Preprint IAS Print-75-QS70, Princeton] Lett. Math. Phys.,\
\textbf{1}(1976), 289--291.

\bibitem[Fd2]{Fd2} Faddeev L.: Einstein and several contemporary tendencies in the theory of elementary particles, in \textit{Relativity, Quanta and Cosmology}, vol. 1, 247--266, Johnson Reprint Co., New York, 1979.

\bibitem[Fd3]{Fd3} Faddeev L.: Knotted solitons. in \textit{Proceedings of the International Congress of Mathematicians}, vol. I (Beijing, 2002),  235--244, Higher Ed. Press, Beijing, 2002.

\bibitem[FN1]{FN1} Faddeev L., Niemi A.: Stable knot-like structures in classical field theory
Nature,\ \textbf{387}(1997), 58--61.

\bibitem[FN2]{FN2} Faddeev L., Niemi A.: Partial duality in $SU(N)$ Yang-Mills theory. Phys. Lett. B,\ \textbf{387}(1999), 214--222.

\bibitem[FH]{FH} Freedman M., He Z.: Divergence-free fields: energy and asymptotic crossing number.
Ann. of Math.(2),\ \textbf{134}(1991), no. 1, 189--229.

\bibitem[GMS1]{GMS1} Giaquinta M., Modica G., Sou\v cek J.: Cartesian currents and variational problems for mappings into spheres. Ann. Scuola Norm. Sup. Pisa Cl. Sci. (4),\ \textbf{16}(1989), no. 3, 393--485.

\bibitem[HL1]{HL1} Hang F., Lin F.:  Topology of Sobolev mappings.  Math. Res. Lett.,\ \textbf{8}(2001),  no. 3, 321--330.

\bibitem[HL2]{HL2} Hang F., Lin F.:  Topology of Sobolev mappings, II.  Acta Math.,\ \textbf{191}(2003),  no. 1, 55--107.

\bibitem[Hl]{Hl} Helgason S.: Differential geometry and symmetric spaces.
Pure and Applied Mathematics, vol. XII, Academic Press, New
York-London, 1962.

\bibitem[HZ]{HZ} Hill C., Zachos C.: Dimensional deconstruction and Wess-Zumino-Witten terms. Phys. Rev. D (3),\ \textbf{71}(2005), no. 4, 046002, 14 pp. 

\bibitem[Hus]{Hus} Husemoller D.: Fibre bundles. Graduate Texts in Mathematics,\ \textbf{20}, Springer-Verlag, New York, 1994. 

\bibitem[IV]{IV} Iwaniec T., Verde A.:  A study of Jacobians in Hardy-Orlicz spaces.
Proc. Roy. Soc. Edinburgh Sect. A,\ \textbf{129}(1999), no. 3,
539--570.

\bibitem[KV]{KV} Kapitanski L., Vakulenko A.: Stability of solitons in $S\sp{2}$ of a nonlinear $\sigma $-model. Dokl. Akad. Nauk SSSR,\ \textbf{246}(1979), no. 4, 840--842. (Russian)

\bibitem[KN]{KN} Kobayashi S., Nomizu K.: Foundations of differential geometry, vol. I, II.
Wiley Classics Library. A Wiley-Interscience Publication, John Wiley
\& Sons, Inc., New York, 1996.

\bibitem[K]{K} Koshkin S.: Homotopy classification of maps into homogeneous spaces. mathGT/0808.0024v1 (submitted to  J. Homotopy Relat. Struct.)

\bibitem[LY2]{LY2} Lin F.,Yang Y.:  Existence of energy minimizers as stable knotted solitons in the Faddeev model. Comm. Math. Phys.,\ \textbf{249}(2004),  no. 2, 273--303.

\bibitem[MS]{MS} Manton N., Sutcliffe P.:  Topological solitons. Cambridge University Press, Cambridge, 2004.

\bibitem[MM]{MM} Marathe K., Martucci G.: The mathematical foundations of gauge theories.
Studies in Mathematical Physics,\ \textbf{5}, North-Holland
Publishing Co., Amsterdam, 1992.

\bibitem[NSK]{NSK} Nawa K., Suganuma H., Kojo T.: Baryons in holographic QCD. Phys. Rev. D,\ \textbf{75}(2007), 086003, 24pp.

\bibitem[Pl]{Pl} Palais R.: Foundations of global non-linear analysis.
W. A. Benjamin, Inc., New York-Amsterdam, 1968.

\bibitem[RRT]{RRT} Robbin J., Rogers R., Temple B.: On weak continuity and the Hodge decomposition. Trans. Amer. Math. Soc.,\ \textbf{303}(1987), no. 2, 609--618.

\bibitem[Sh2]{Sh2} Shabanov S.: On a low energy bound in a class of chiral field theories with solitons.  J. Math. Phys.,\ \textbf{43}(2002),  no. 8, 4127--4134

\bibitem[St]{St} Steenrod N.: The topology of fibre bundles. Princeton Landmarks in Mathematics, Princeton Paperbacks, Princeton University Press, Princeton, NJ, 1999.

\bibitem[Th]{Th} Thaler J.: Little technicolor. JHEP, (2005), no. 07, 024, 15 pp.

\bibitem[Ul1]{Ul1} Uhlenbeck K.: Connections with $L\sp{p}$ bounds on curvature.
Comm. Math. Phys.,\ \textbf{83}(1982), no. 1, 31--42.

\bibitem[We]{We} Wehrheim K.: Uhlenbeck compactness. EMS Series of Lectures in Mathematics, EMS, Z\"urich, 2004.

\bibitem[Wh]{Wh} White B.: Homotopy classes in Sobolev spaces and the existence of energy minimizing maps. Acta Math.,\ \textbf{160}(1988), no. 1-2, 1--17.

\end{thebibliography}

}

\end{document}